\documentclass[12pt]{iopart}
\usepackage{iopams,graphicx,color}
\usepackage[utf8]{inputenc}
\usepackage{epstopdf}

\newcommand{\F}{\mathrm{f}}
\newcommand{\B}{\mathrm{b}}
\newcommand{\U}{\mathrm{u}}
\begin{document}

\title[Kinetic proofreading]{Thermodynamics of accuracy in kinetic proofreading: Dissipation and efficiency trade-offs}

\author{Riccardo Rao$^{1}$\footnote[1]{To whom correspondence should be
addressed. Present address: Complex Systems and Statistical Mechanics, University of Luxembourg, L--1511 Luxembourg,
G.D. Luxembourg}, and
Luca Peliti$^{2}$}

\address{$^{1}$ Dipartimento di Fisica, Università ``Federico II'', Complesso Monte S.~Angelo, I--80126 Napoli, Italy}

\address{$^{2}$ Simons Center for Systems Biology, Institute for Advanced Study, Einstein Drive, Princeton NJ 08540, USA}

\eads{\mailto{riccardo.rao@uni.lu}, \mailto{luca@peliti.org}}

\begin{abstract}
The high accuracy exhibited by biological information transcription processes is due to kinetic proofreading, i.e., by a mechanism which reduces the error rate of the information-handling process by driving it out of equilibrium. 
We provide a consistent thermodynamic description of enzyme-assisted assembly processes involving competing substrates, in a Master Equation framework.
We introduce and evaluate a measure of the efficiency based on rigorous non-equilibrium inequalities. 
The performance of several proofreading models are thus analyzed and the related time, dissipation and efficiency vs.\ error trade-offs exhibited for different discrimination regimes.
We finally introduce and analyze in the same framework a simple model which takes into account correlations between consecutive enzyme-assisted assembly steps.
This work highlights the relevance of the distinction between energetic and kinetic discrimination regimes in enzyme-substrate interactions.
\end{abstract}

\pacs{%
87.10.Vg (Biological information),
05.70.Ln (Nonequilibrium and irreversible thermodynamics),
87.18.Tt (Noise in biological systems)
}


\maketitle
\section{Introduction}
\label{sec:Introduction}
Life is a non-equilibrium process involving the handling of information, both internal (about the organism) and external (about the environment). It comes therefore as no surprise that the mechanisms of information processing in living beings are highly sophisticated. About four decades ago it was recognized by both Jacques Ninio~\cite{ninio:proofreading} and~J.J.~Hopfield~\cite{hopfield:proofreading} that nucleic acid transcription and translation could achieve unexpected fidelity by inserting a strongly irreversible step in the process, a mechanism that was given the name of \textit{kinetic proofreading}. The necessity of an irreversible step in error correction is in agreement with the principle stated by Rolf Landauer~\cite{landauer:irreversibility}, that any \textit{logically} irreversible manipulation of information must be accompanied by a corresponding entropy increase either in the system itself or in the environment (cf.~\cite{bennett:notes}). The validity of this principle far from equilibrium has been recently confirmed by M.~Esposito and~C.~van~den~Broeck~\cite{esposito:farfrom} on the basis of stochastic thermodynamics. Hopfield and~collaborators were able to collect evidence of kinetic proofreading mechanisms in tRNA aminoacylation~\cite{hopfield:evidence,yamane:evidence}. A different scheme in which high fidelity can be achieved in these fundamental molecular processes was later suggested by C.~Bennett~\cite{bennett:proofreading}. One may therefore argue that one of the fundamental requirements of living processes---the need for an enhanced fidelity---necessarily implies that these processes take place far from thermodynamical equilibrium. The subtle thermodynamics involved in copying a biopolymer on a template has been more recently investigated by D.~Andrieux and~P.~Gaspard~\cite{andrieux:randomness,andrieux:copol08,andrieux:erasure} as well as by~J.~R.~Arias-Gonzalez~\cite{arias:DNAentropy}. Detailed kinetic models of the DNA transcription process have been analyzed, from the point of view of both their thermodynamic efficiency and their fidelity, by M.~Voliotis and collaborators~\cite{voliotis:DNAtranscription}. A more general model, involving several reaction loops, has been thoroughly investigated by A.~Murugan, D.~Huse and~S.~Leibler~\cite{murugan:speed}. This work has pointed out that the Hopfield model can work with high accuracy and speed in a special regime which had remained unnoticed in the original work. This analysis has led to identify some rather general design principle for proofreading networks and to point out that a network with a given architecture (with given reaction constants) can work in several regimes, switching from one to another by just changing the distance from equilibrium (e.g., by manipulating the ATP/ADP imbalance)~\cite{murugan:regimes}.

Our understanding of the features which determine the performance of proofreading networks has been improved by the work of~P.~Sartori and~S.~Pigolotti~\cite{sartori:discrimination}.
These authors emphasize the difference between \textit{energetic} and~\textit{kinetic} discrimination regimes, according to the mechanism which leads the enzyme to discriminate between substrates.
Energetic discrimination relies on a difference in substrate-enzyme affinity, while kinetic discrimination relies on a difference in the base reaction rate.
In the simple copolymerization scheme introduced by C.~Bennett in \cite{bennett:proofreading}, the former becomes more precise when the reaction takes place closer to equilibrium, while the latter becomes more precise when the reaction takes place faster.
Thus the difference between the Ninio and~Hopfield scheme on the one side, and the Bennett~one on the other side, appears to be mainly due to the fact that the first one works in the energetic regime, and the other in the kinetic one.

The insights gained by recent progress in non-equilibrium statistical physics and stochastic thermodynamics on the thermodynamics of information have been recently reviewed in~\cite{parrondo:thermodynamics}. In~\cite{celani:harvesting} the thermodynamics of information has been exploited to set limits to the amount of information that can be stored in the memory of a stochastic measurement device given an energy budget. An efficiency measure for information handling in cellular sensing was proposed in~\cite{seifert:efficiency}, and the analogy between sensing and kinetic proofreading has been pointed out and discussed in~\cite{seifert:analogy}.

In the present work we introduce a thermodynamic description for enzyme-assisted assembly processes embedding kinetic proofreading.
We thus reconsider a number of simple proofreading schemes in the light of these recent developments, and especially of the classification of discrimination regimes proposed by Sartori and~Pigolotti.
We are able to introduce a measure of the efficiency of the process, which is related to the entropy production, and discuss how the error rate, the reaction speed and the efficiency vary, for a given architecture, as a function of one another. We also consider in a simple setting the effects of error correlations in polymer synthesis on a template.
	
The general reaction scheme for enzyme-assisted assembly with proofreading is described in sec.~\ref{sec:description}.
In the same section we provide the kinetic and thermodynamic description, and all the relevant quantities are defined.
The simplest reaction scheme, the Michaelis-Menten (MM) model (without proofreading), is defined and discussed in sec.~\ref{sec:Michaelis-Menten}. The ground-breaking model introduced by Ninio and~Hopfield is discussed in sec.~\ref{sec:Hopfield}.
The more general scheme introduced by Murugan, Huse and~Leibler is taken into account in sec.~\ref{sec:Murugan}.
In all these models, the enzyme assisted reaction is treated as a renewal process, in that when the product is released, the enzyme goes back to its neutral, ``bare'', state. In order to make a first step in understanding the role of correlations in this kind of process, we introduce and discuss in sec.~\ref{sec:correlations} a simple model in which the effect of a correct or wrong incorporation affects the rate constant of the next reaction. We close the work with some conclusions and hints for further investigations.
\section{Description}
\label{sec:description} 
Enzyme-catalyzed chemical reaction networks involving competing substrates represent the typical scheme appearing in transcription processes. In this framework, the enzyme catalyzes the transcription reactions for one of the competing substrates, e.g., for one activated nucleotide in DNA replication or transcription. More generally, one can consider an enzyme-catalyzed assembly reaction like, e.g., tRNA aminoacylation, in which the tRNA-enzyme complex captures an activated amino acid in the cytoplasm and either binds it to the tRNA or rejects it. However, only one of the substrate carries on the correct information and leads to the correct final product. The question which thus arises is how the enzyme can reduce the rate of errors for a given energy consumption.

\begin{figure}
\centering
	\includegraphics[width=.7\textwidth]{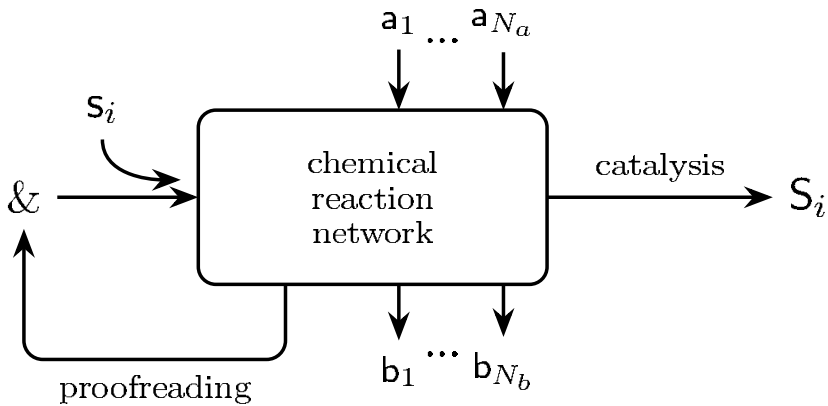}
	\caption{General scheme of an enzyme-assisted assembly process.}
\label{fig:generic}
\end{figure}

The general scheme of an enzyme-catalyzed assembly is described in~fig.~\ref{fig:generic}. The enzyme, in its free state $\mathsf{\&}$, reacts with one of the competing substrates $\mathsf{s}_i$, thus forming an activated complex. The chemical enzyme-substrate complex then undergoes a series of chemical reactions, which represent the intermediate stages needed to complete the assembly process. During these stages the enzyme performs the proofreading, i.e., it can dissociate from the substrate by means of the proofreading pathway. This is more likely to happen if the substrate is the wrong one. Otherwise, the catalysis transition terminates the step, leading the system to the final product state $\mathsf{S}_i$ (corresponding to the substrate $\mathsf{s}_i$), and restoring the free enzyme state $\mathsf{\&}$. Let us remark that the intermediate reactions could also involve further chemical species, denoted in fig.~\ref{fig:generic} by $\left\lbrace \mathsf{a}_i \right\rbrace$ and $\left\lbrace \mathsf{b}_i \right\rbrace$. They can either be part of the final product state or provide the system with the chemical energy needed to carry on the processes (one example is ATP hydrolysis \cite[p.~80-87]{alberts:cell5}). In our framework, we treat these species as chemicals whose concentrations are kept constant by the environment. This assumption simplifies the description, in that the only species whose concentration varies in time is the enzyme-substrate complex.
Also, the chemical network can thus be correctly represented by a graph---the proper way to represent generic chemical networks is by means of hyper-graphs~\cite{klamt:hypergraphs,polettini:networks1}.
In this way, the nodes of the graph represent the metastable states of the complex, and the reactions the oriented arcs which connect these states~\cite{hill:biochemical,schnakenberg:network}.

We shall call \textit{internal} all the reaction steps of the process (including the proofreading one), except the final catalysis step. Internal reactions are assumed to be reversible and their rates can be expressed in the Kramers form:
\begin{equation}\label{eq:Kramers}
 	k = \Omega\, \rme^{-\Delta},
\end{equation}
where $\Omega$ is a constant characterizing the reaction pathway (i.e., it is equal for the inverse reaction), while $\Delta$ takes into account both the reaction activation energy and the concentrations of the external chemical species. (We assume that all reactions take place in an environment with a constant absolute temperature $T$, and we set $k_{\mathrm{B}}T\equiv 1$ from now on.) In this way, the rates determine the non-equilibrium conditions and local detailed balance is preserved.

The final catalysis transition terminates the assembly process.
Biologically, it corresponds to the stage beyond of which the enzyme does not process that substrate anymore.
Thus, the final product states $\mathsf{S}_i$ can be regarded as absorption states of the process, where the free-enzyme state is restored and a new assembly process can start again. We can thus describe the occurrence of many instances of the process, one after the other, by disregarding their final product state and by considering this reaction as leading irreversibly to the free-enzyme state (cf.~\cite[\S~8]{hill:biochemical}).

This description is tacitly based on the assumption that the steps are similar and independent, and that there is no correlation between successive steps. This is not always the case.
For example, in transcription processes involving a template strand, like DNA replication and RNA transcription, the secondary structure of the strands imposes specific correlations between the monomers~\cite{santalucia:DNAmotifs}.
These correlations, in turn, imply that the strand configuration plays a (fundamental) role in both the individuation and the proofreading of errors \cite{johnson:mismatch}. 

We can take into account these correlations in two ways:
\begin{itemize}
\item By considering multiple assembly processes, each of which depends on the result of the previous step.
	The sequentiality is obtained by means of the final catalysis transitions.
	An example is provided in sec.~\ref{sec:correlations}.
\item By considering an internal chemical network in which the enzyme reacts with multiple substrates.
	This means that the internal network encompasses several assembly stages.
	In this case, the final catalysis basically corresponds to the assembly of the first substrate in the queue, while the following ones are still being processed.
	The correlations between successive assembly stages are thus embedded in the internal chemical network.
\end{itemize}
 
The second kind of correlations appears to be similar to that introduced by J.J.~Hopfield in \cite{hopfield80}.
In this scheme, when the enzyme proofreads the substrate, it accesses to a lower free-energy enzyme state.
From this lower energy state, the reaction continues without proofreading, but when the enzyme catalyses the final reaction, the high-energy state is restored.
Indeed, the final reaction is supposed to provide the enzyme with the energy required by this restoration (hence the name ``relay proofreading'').
Finally, an effective reduction of the error rate can be achieved without increasing the energetic discrimination constant by properly tuning the rate constants.

Once formalized the basic assumptions about the kinetics of our system, we need to describe how the enzyme can discriminate the substrates.
Since it undergoes the same kind of reactions with any substrate, the structure of reaction network leading from the free enzyme state to the final product state will be the same.
In other words, the network topology does not discriminate.
According to~\cite{sartori:discrimination}, the substrates discrimination is embedded in the rate constants.
Indeed, the substrates are distinguished either because of differences of chemical complex binding energies (the typical example is DNA replication and RNA transcription, in which there is an affinity difference between matching and mismatching nucleotide pairs), or because the activation energies related to the reaction involving the right substrate are lower than for the other ones (figure~\ref{fig:mm}(left), describing the energy landscape of the Michaelis-Menten reaction scheme, can help in visualizing these ideas).
Following~\cite{sartori:discrimination}, we will refer to the first type of discrimination as \emph{energetic} while the second one as \emph{kinetic}.
It is worth noting that all of these differences appear in the rate constants.

The rates of the final catalysis transitions will not be written in Kramers form, due to their irreversible nature. We shall however assume that these rates are the same for all the substrates, since there is no substrate discrimination in this step.
This assumption relies on the fact that most often the region of the enzyme involved in the discrimination is located far from that involved in the final binding, cf.~\cite[p.~224]{bialek:biophysics}.

The stochastic description of the process is described by a Master Equation \cite{schnakenberg:network}, \cite[Ch.~5]{kampen:stochastic}.
Denoting by $p_{i}(t)$ the probability of finding the chemical complex in the state $i$ at time $t$,
we can write it as
\begin{equation}
\label{eq:me} 
	\dot{p}_{i}(t) = \sum_{j\,(\neq i)} \left[k_{ij} p_{j}(t) - k_{ji} p_{i}(t)\right] ,
\end{equation}
where $k_{ij}$ denotes the rate constant regulating the probability rate of the reaction leading from the state $j$ to the state $i$.

As shown in~\cite{schnakenberg:network} and~\cite[Ch.~5]{kampen:stochastic}, this kind of processes evolves towards a unique \emph{steady state} in which the probabilities are constant. Since we are not interested in the complete time evolution of these processes, but only in the steady-state regime of operation, we shall not describe in the following the transient initial stochastic dynamics. We shall instead directly focus on the steady state, which can be evaluated by either solving the linear set of equation related to the master equation (\ref{eq:me}) or by means of graphical methods~\cite[\S~6]{hill:biochemical}, \cite[\S III]{schnakenberg:network}. Given the steady-state probabilities~$(\bar{p}_{i})$, which are functions of the rate constants, we can evaluate all the relevant kinetic and thermodynamic quantities. Specifically, for any given reaction scheme, the kinetic quantities we are interested in are described below, while the thermodynamic ones are discussed in the next paragraph.
\begin{description}
	\item[The error rate:] \ 
		\begin{equation}
		\label{def:er} 
	\xi \equiv \frac{\mbox{rate of wrong catalysis: }J_{\mathsf{w}}}{\mbox{total rate of catalysis: }J_{\mathsf{w}} + J_{\mathsf{r}}},
		\end{equation}
		which describes the average rate of wrong outcome observed;
	\item[The mean step time:] \ 
	\begin{equation}
	\label{def:mst} 
		\tau \equiv \frac{1}{\mbox{total rate of catalysis: }J_{\mathsf{w}} + J_{\mathsf{r}}} ,
	\end{equation}
	describing the average time needed to complete a step
	(this definition follows from that of mean first passage time of absorption in the final product state \cite[Ch.~8]{hill:biochemical} \cite{redner:first-passage};
\end{description}

\subsection*{Thermodynamic description}

At a first glance, the thermodynamic description of proofreading processes as described so far is complicated by the presence of the irreversible catalysis.
However, it is clear that these transitions merely describe the succession of the enzymatic assembly steps and do not take part in the discrimination itself.
Thus we shall not take into account these steps in the expression of the entropy production and entropy flow.
Following \cite{hill:biochemical,schnakenberg:network,prigogine:thermodynamics} the entropy production and entropy flow in the process are respectively given by
\begin{eqnarray}
\label{def:ep}
	\dot{S}_{\rmi} &\equiv \frac{1}{2} {\sum_{i,j}}'
		\underbrace{\left[ k_{ij} p_j - k_{ji} p_i \right]}_{\mbox{probability flux}}
		\underbrace{\ln \frac{k_{ij} p_j}{k_{ji} p_i}}_{\mbox{affinity}} ,\\
\label{def:ef}
	\dot{S}_{\rme} &\equiv - \frac{1}{2} {\sum_{i,j}}'
		\left[ k_{ij} p_j - k_{ji} p_i \right]
		\ln \frac{k_{ij}}{k_{ji}}.
\end{eqnarray}
The primed sums run over all pairs of states, but exclude the final catalysis transition.
Note that we consider the entropy flow as positive if it enters the system.
The time derivative of the system's Gibbs entropy is given by
\begin{equation}
\label{eq:balance}
-\frac{\rmd}{\rmd t} {\sum_{i}}p_{i}(t) \ln p_{i}(t) = \dot{S}_{\rmi} + \dot{S}_{\rme} - \dot{S}_{\mathrm{c}},
\end{equation}
where the last term takes into account the contribution to the entropy change due to the catalysis transitions. More precisely, $\dot{S}_{\mathrm{c}}$ can be viewed as the entropy change in the environment (hence the minus sign) due to the completion of the assembly.
At the steady state, the expression on the left-hand side vanishes, but the single terms on the right-hand side do not vanish in general.
By multiplying the entropy rates in the steady state by the mean-step time $\tau$ (\ref{def:mst}) we obtain respectively:
\begin{description}
\item[The entropy production per step:] \ 
	\begin{equation}
	\label{def:epps}
		\Delta_{\rmi} S \equiv \tau \dot{S}_{\rmi}
	\end{equation}
which is proportional to the amount of free energy irreversibly lost during a step;
\item[The entropy flow per step:] \ 
	\begin{equation}
	\label{def:efps} 
		\Delta_{\rme} S \equiv \tau \dot{S}_{\rme}
	\end{equation}
which is the entropy reversibly exchanged with the reservoirs during a step.
Let us highlight that in all the reaction schemes we describe, the process evolves from the free-enzyme state to lower-energy bound-substrate states.
Thus these processes are exothermic and $\Delta_{\rme} S$ is negative.
\item[The entropy change per step:] \ 
	\begin{equation}
	\label{def:ecps} 
		\Delta_{\mathrm{c}} S \equiv \tau \dot{S}_{\mathrm{c}}
	\end{equation}
	which is the entropy change in the environment due to the completion of just one step.
Since assembly mechanisms bind free monomers into $\mathsf{W}$ and $\mathsf{R}$ final products, thus decreasing their available phase space, $\Delta_{\mathrm{c}} S$ is negative. Furthermore, the fewer errors occur, the more ordered is the final state, and the smaller is $\Delta_{\mathrm{c}} S$.
The explicit expression of this term depends on the chemical network topology.
In sec.~\ref{sec:Michaelis-Menten}, we provide it for assembly processes devoid of correlations and with two competing substrates (eq.~\ref{expr:mm:ec}).
\end{description}

The above-described quantities suggest us to define the efficiency of the proofreading process in the following way
\begin{equation}
\label{def:effps} 
	\eta \equiv\frac{\Delta_{\mathrm{c}} S}{\Delta_{\rme} S} = 1 + \frac{\Delta_{\rmi} S}{\Delta_{\rme} S} .
\end{equation}
Since both $\Delta_{\mathrm{c}} S$ and $\Delta_{\rme}S$ are negative and $\Delta_{\rmi}S$ is positive, this expression yields a value lying between 0 and 1.
(An analogous definition of efficiency can be found in~\cite{diana:erasing}, applied to the process of information erasing in fermionic bits.)

We shall exploit this thermodynamic description in the following sections.
In particular, we will recover the thermodynamic trade-offs obtained in copolymerization models \cite{bennett:proofreading, andrieux:copol08, andrieux:copol09, sartori:discrimination}, and we will evaluate the corresponding efficiency.
Finally, we highlight the following two points: 
	(i) the splitting of the network into \emph{internal} and \emph{external} steps resembles the splitting into \emph{observable} and \emph{masked} steps, whose thermodynamics has been analysed in \cite{shiraishi:masked};
	(ii) even if we had taken into account the irreversible catalysis path in the thermodynamic description, its contribution to the entropy production would have been finite \cite{murashita14}.

\section{The Michaelis-Menten model}
\label{sec:Michaelis-Menten}
\begin{figure}
\centering
	\includegraphics[width=.45\textwidth]{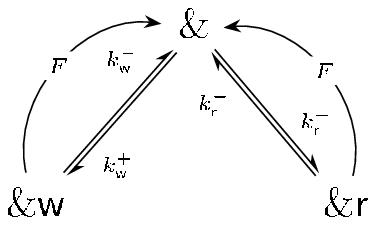} \quad
	\includegraphics[width=.45\textwidth]{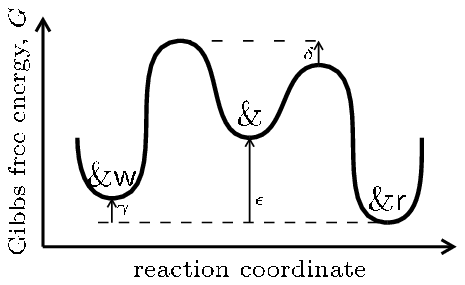}
\caption{
	Michaelis-Menten (MM) reaction scheme (left) and enthalpy landscape (right) for enzyme-assisted assembly processes involving two competing substrates.
	As mentioned in sec.~\ref{sec:description}, the catalysis transitions are directly and irreversibly connected to the free-enzyme state.
	The driving energy, $\epsilon$, represents the enthalpy difference between the free enzyme state and the right product.
	The difference in binding energies between the wrong and right activated complex, $\gamma$, provides the \emph{energetic discrimination}, whereas the difference of activation energies, $\delta$, provides the \emph{kinetic discrimination}.
	Let us emphasize that our representations of the free enthalpy landscapes simply provide a visual aid for understanding the models, and are not to be interpreted as experimentally reproducible curves.
}
\label{fig:mm}
\end{figure}

Here we summarize the minimal reaction scheme describing assembly mechanisms, which follows the \emph{Michaelis-Menten} (MM) enzymatic kinetics \cite[sec.~4.5]{bialek:biophysics} \cite{hopfield:proofreading}, fig.~\ref{fig:mm}(left).

In this simple reaction scheme, the enzyme, in its free enzyme state $\mathsf{\&}$, reacts with one of the two substrates, $\mathsf{r}$ or $\mathsf{w}$, thus forming the activated complex, $\mathsf{\&} \mathsf{r}$ or $\mathsf{\&} \mathsf{w}$, respectively.
Subsequently, it concludes the assembly step by means of the final catalysis transition.
The MM reaction scheme provides us with a starting point for introducing the kinetic and thermodynamic description of the discrimination process.
The present analysis essentially reproduces that of~\cite{sartori:discrimination}, in a form that can be easily extended to more complicated networks.

Together with the reaction scheme shown in fig.~\ref{fig:mm}(left), we consider the free-energy landscape in fig.~\ref{fig:mm}(right). Given this energetic scheme and using the Kramers form of the rate constants we can write the reaction rates in the following form
\begin{equation}\label{expr:mm}
\eqalign{
	 	k_{\mathsf{r}}^{+} = \omega  \rme^{\delta + \epsilon},\qquad	& k_{\mathsf{r}}^{-} =  \omega \rme^{\delta }, \\ 
	 	k_{\mathsf{w}}^{+} = \omega \rme^\epsilon, 			& k_{\mathsf{w}}^{-} = \omega \rme^{\gamma },}
\end{equation}
where $\omega$ is the overall time scale of the reaction pathway and $\epsilon$ denotes the driving force acting on it. The \emph{discrimination constants}, $\gamma$ and $\delta$, quantify the difference of binding free energy between the correct and the wrong activated complex and the difference of activation free energy between the correct and wrong free-enzyme-to-activated-complex reaction, respectively.

The master equation for this system reads
\begin{equation}
\eqalign{
 		\dot{p}_{\mathsf{\&}} = \left(F + k_{\mathsf{w}}^- \right) p_{\mathsf{w}} + \left( F + k_r^-\right)  p_{\mathsf{r}} - \left(k_{\mathsf{w}}^+ + k_{\mathsf{r}}^+\right) p_{\mathsf{\&}} ,\\
		\dot{p}_{\mathsf{w}} = k_{\mathsf{w}}^+ p_{\mathsf{\&}} - \left( F + k_{\mathsf{w}}^- \right) p_{\mathsf{w}}, \\
		\dot{p}_{\mathsf{r}} = k_{\mathsf{r}}^+ p_{\mathsf{\&}} -  \left(F + k_{\mathsf{r}}^-\right)  p_{\mathsf{r}} ,
 	} 	\end{equation}
which, solved in the steady-state regime, leads to the following expressions for the probabilities:
\begin{equation}
 	\eqalign{
		\bar{p}_{\mathsf{\&}} = \frac{1}{\Sigma} \left[ F^2 + F  \left(k_{\mathsf{r}}^- + k_{\mathsf{w}}^-\right) + k_{\mathsf{r}}^- k_{\mathsf{w}}^- \right], \\
 		\bar{p}_{\mathsf{w}}  = \frac{1}{\Sigma} k_{\mathsf{w}}^+  \left(F + k_{\mathsf{r}}^- \right), \\
 		\bar{p}_{\mathsf{r}}  = \frac{1}{\Sigma} k_{\mathsf{r}}^+  \left(F + k_{\mathsf{w}}^- \right).}
 	\end{equation}
Here we have defined the normalization factor $\Sigma$ by
\begin{equation}	
\Sigma \equiv F^2 + F(k_{\mathsf{r}}^- +k_{\mathsf{r}}^++k_{\mathsf{w}}^- +k_{\mathsf{w}}^+ )+k_{\mathsf{r}}^-  (k_{\mathsf{w}}^- +k_{\mathsf{w}}^+ )+k_{\mathsf{r}}^+k_{\mathsf{w}}^- .
\end{equation}

Following the definition given in (\ref{def:er}), and the Kramers form for the rate constants~(\ref{expr:mm}), we express the error rate as follows:
\begin{equation}
\label{expr:mm:er}
	\xi \equiv \frac{F \bar{p}_{\mathsf{w}}}{F \bar{p}_{\mathsf{w}} + F \bar{p}_{\mathsf{r}}} = \frac{\rme^{\delta } \omega +F}{\left(\rme^{\gamma }+1\right) \rme^{\delta } \omega
   +\left(\rme^{\delta }+1\right) F}.
\end{equation}
Analogously, the mean step time defined in (\ref{def:mst}) as
\begin{equation}
\label{expr:mm:mst} 
	\tau \equiv \frac{1}{F \bar{p}_{\mathsf{w}} + F \bar{p}_{\mathsf{r}}}, 
\end{equation}
while the entropy production per step and the efficiency are obtained by a simple application of (\ref{def:epps}) and~(\ref{def:effps}).
Finally, the entropy change per step is expressed by
\begin{equation}
\label{expr:mm:ec}
\Delta_{\mathrm{c}} S = - \frac{1}{\bar{p}_{\mathsf{r}} + \bar{p}_{\mathsf{w}}} \left[ \bar{p}_{\mathsf{r}} \ln \frac{\bar{p}_{\mathsf{r}}}{\bar{p}_{\mathsf{\&}}} + \bar{p}_{\mathsf{w}} \ln \frac{\bar{p}_{\mathsf{w}}}{\bar{p}_{\mathsf{\&}}} \right].
\end{equation}

\begin{figure}
\centering
	\begin{tabular}{cc}
		\includegraphics[width=.47\textwidth]{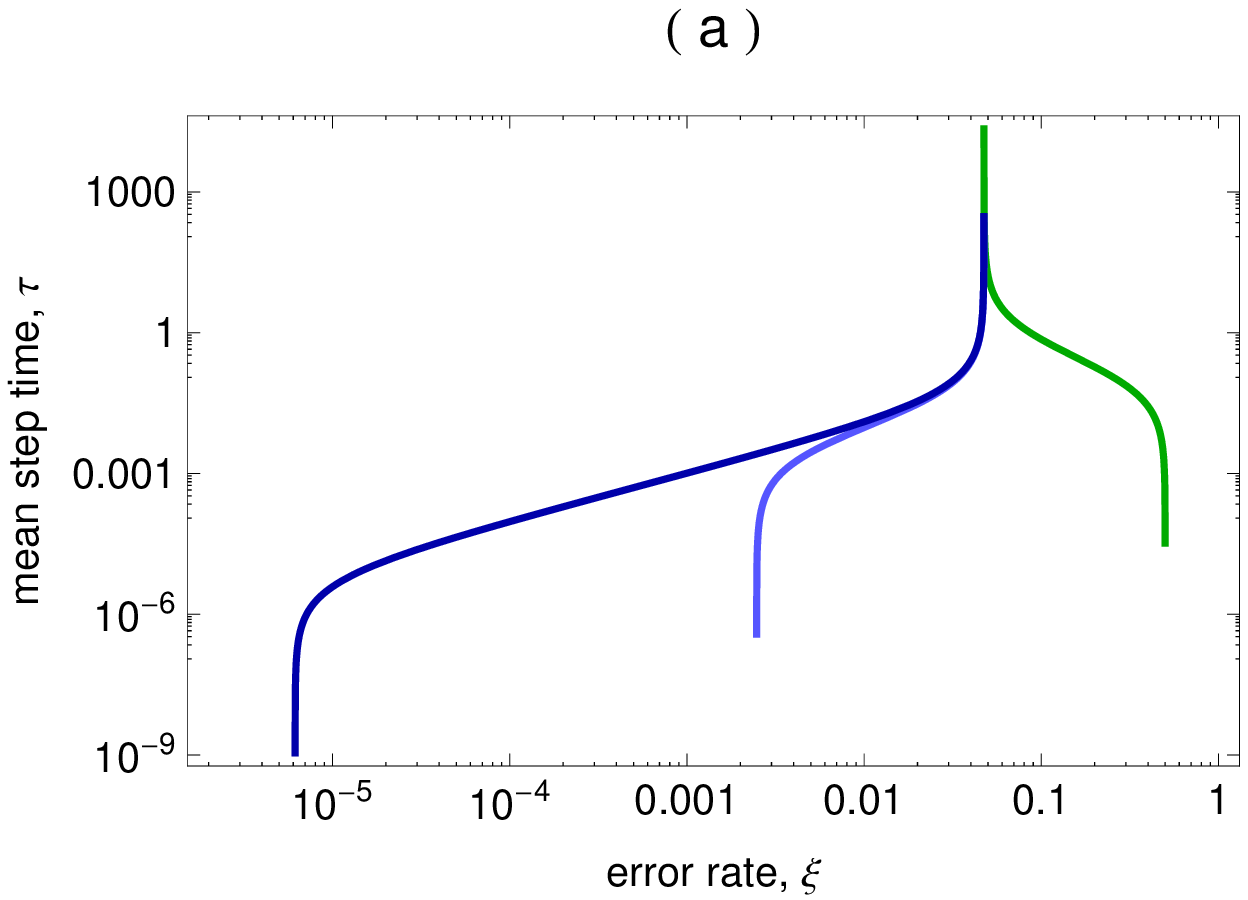} &
		\includegraphics[width=.47\textwidth]{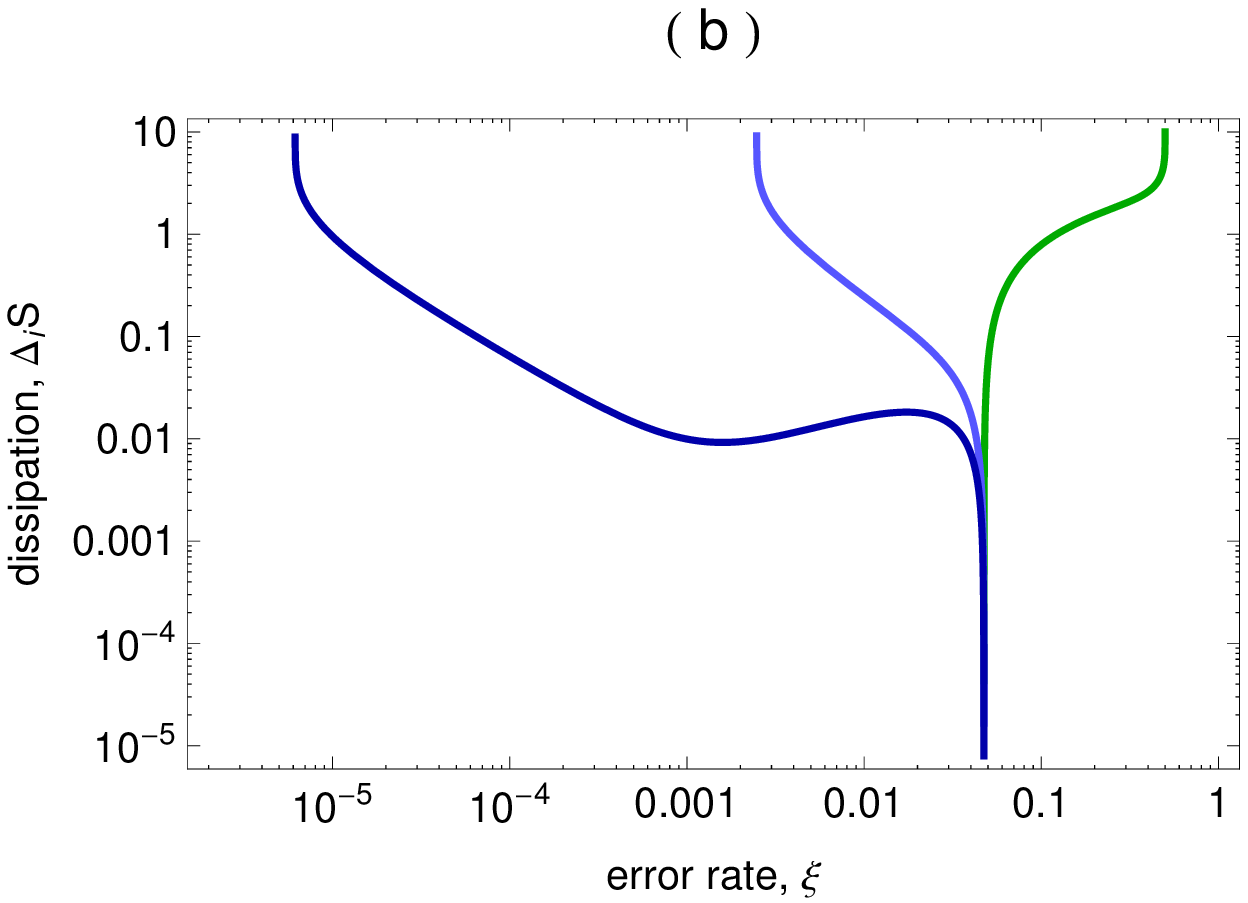} \\
		\includegraphics[width=.47\textwidth]{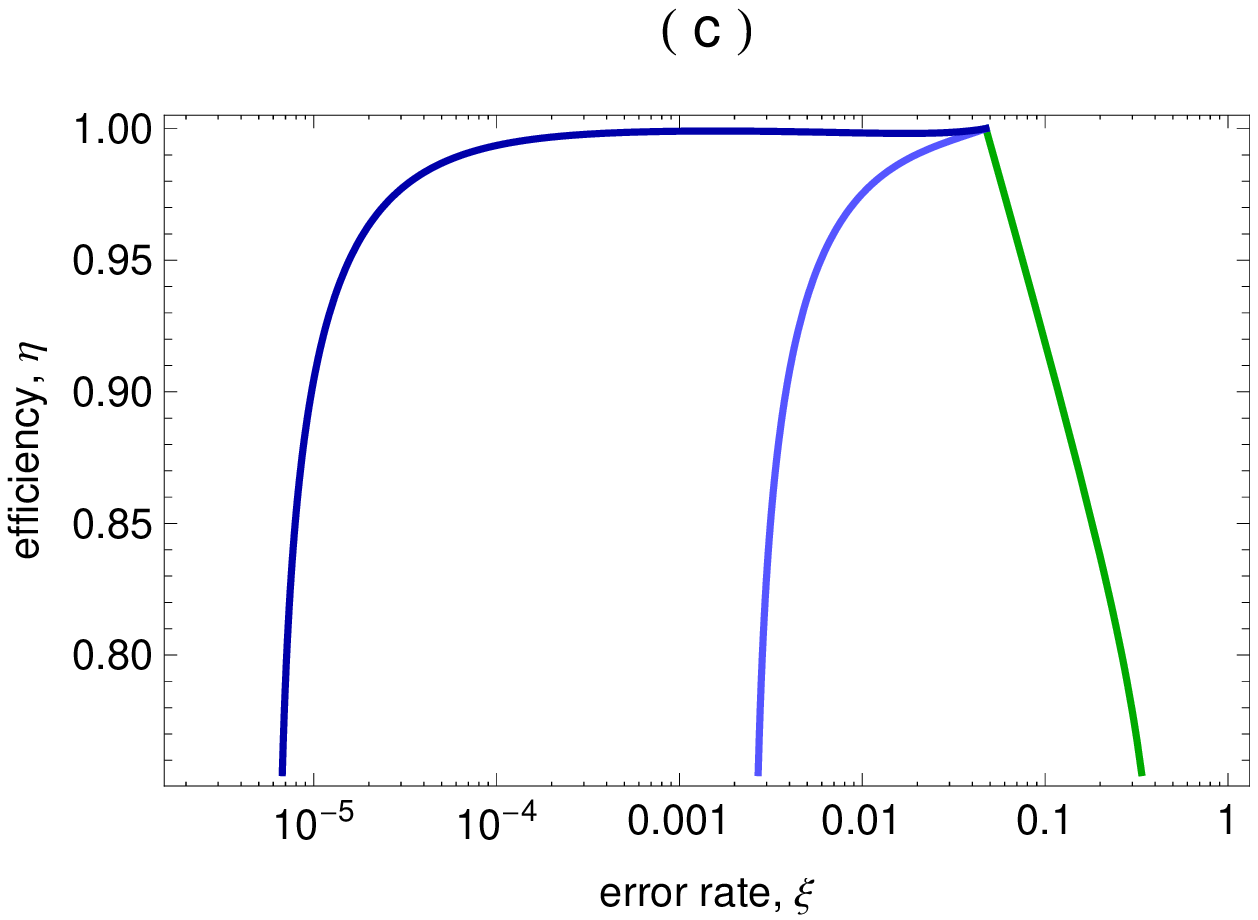} &
		\includegraphics[width=.47\textwidth]{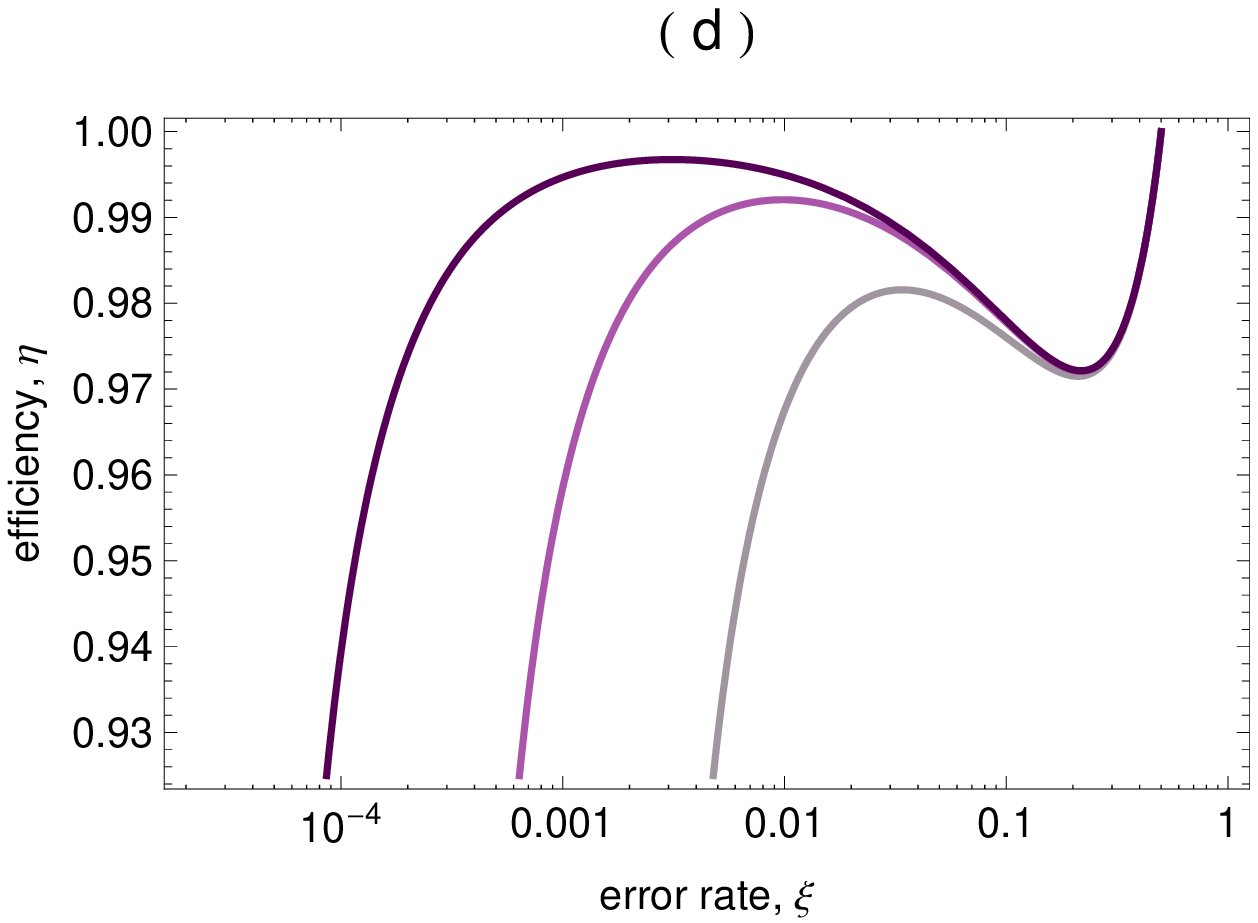}
	\end{tabular}
\caption{
	Mean step time (a), dissipation per step (b) and efficiency (c) versus the error rate in the MM model for different discrimination regimes.
The reported graphs correspond to the following values for the coefficients: $\gamma = 3$, $\epsilon = 10$, $\omega = 1$ and $\delta = 0$ (green, purely energetic discrimination), $\delta = 6$ (light blue, mainly kinetic discrimination), $\delta = 12$ (dark blue, mainly kinetic discrimination).
	(d)~Efficiency-error trade-off in the purely kinetic regime of discrimination. 
	The coefficients values are: $\gamma = 3$, $\epsilon = 10$, $\omega = 1$ and $\delta = 6$ (light purple), $\delta = 8$ (purple), $\delta = 10$ (dark purple).
	Interestingly, the efficiency exhibits a relative maximum in kinetic regime of discrimination and the maximum value tends to one as $\delta$ increases.
}
\label{fig:mm:tos}
\end{figure}

In order to determine the trade-offs between the error rate and the other kinetic and thermodynamic quantities, we observe that the former (eq.~\ref{expr:mm:er}) is a monotonic function of the catalytic rate, $F$.
We can thus invert $\xi(F)$ and substitute the result into the expression of the mean step time, entropy production per step and efficiency.
In this way, all the thermodynamic quantities are expressed in terms of the error rate, which becomes an independent variable.
The time-error, dissipation-error and efficiency-error trade-offs are shown in fig.~\ref{fig:mm:tos}.

In the \emph{energetic discrimination regime} \cite{sartori:discrimination} ($\gamma >\delta$), $\xi$ monotonically increases for increasing catalytic rate.
The minimum error rate is thus achieved in a quasistatic condition, i.e., vanishing catalysis, $F\rightarrow 0$.
In this limit, the mean step time diverges, the dissipation vanishes and the efficiency reaches unity.
The green curves in fig.~\ref{fig:mm:tos} describe the trade-offs in the purely energetic discrimination regime and exhibits the above-described behavior close to the minimum error rate.
On the contrary, in the \emph{kinetic discrimination regime} \cite{sartori:discrimination} ($\delta >\gamma$), $\xi$ is monotonically decreasing with increasing $F$.
In this case, the faster the process, the more discriminating it becomes, but this is accompanied by a larger dissipation needed to keep the process out of equilibrium.
This discrimination regime is described by the blue curves in fig.~\ref{fig:mm:tos}.
The smallest achievable error rate depends on the discrimination regime and has the following expression:
\begin{equation}
	\xi_{\min} = \frac{1}{\rme^{ \max \left\{\delta, \gamma\right\} } + 1 } \simeq \rme^{- \max \left\{ \delta, \gamma \right\} }.
\end{equation}
Since only the maximum between the $\gamma$ and $\delta$ contributes to the minimum error rate, there is no way of enhancing the discriminatory power of the model by combining the discrimination factors.

The time-error and dissipation-error trade-offs are the same of those obtained by copolymerization models \cite{sartori:discrimination}.
However, the definition of efficiency allows us to shed new light on this processes.
Indeed, the efficiency-error trade-off in the kinetic discrimination regime exhibits a maximum in correspondence to the relative minimum of the dissipation.
As the kinetic discrimination constant increases this maximum tends to one, as shown in fig.~\ref{fig:mm:tos}(d).

\section{The Ninio-Hopfield model}
\label{sec:Hopfield}

\begin{figure}
\centering
\includegraphics[width=.7\textwidth]{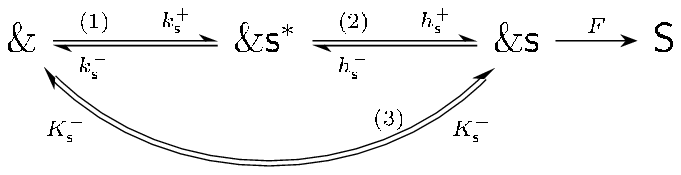}  \\
\includegraphics[width=.7\textwidth]{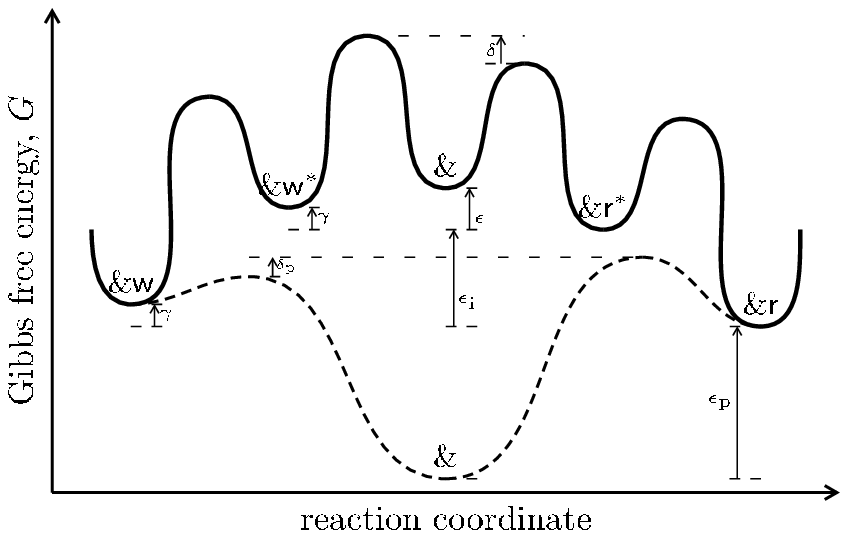} 
\caption{
	(top) Hopfield model reaction scheme; 
	$\mathsf{s}$ denotes the generic substrate, i.e., $\mathsf{s}\in\{\mathsf{r},\mathsf{w}\}$.
	(bottom) Free-energy diagram for the Hopfield model. The thick continuous line in (b) represents the Gibbs free energy potential along the reaction coordinate relative to the first two reactions: $\mathsf{\&} \rightleftharpoons \mathsf{\&} s^{\ast}$ and $\mathsf{\&} s^{\ast} \rightleftharpoons \mathsf{\&} s$. The dashed line corresponds to the proofreading pathway. The energy difference between two consecutive free-enzyme states ($\mathcal{A} \equiv \epsilon + \epsilon_{i} + \epsilon_{\mathrm{p}}$) is the free energy consumed when a cycle $\mathsf{\&} \rightarrow \mathsf{\&} s^{\ast} \rightarrow \mathsf{\&} s \rightarrow \mathsf{\&}$ is performed. The discrimination is performed along the first and/or the proofreading reaction, where $\gamma$, $\delta$ and $\delta_{\mathrm{p}}$ are the discrimination constants.
}
\label{fig:h}
\end{figure}

\begin{figure}
\centering
\begin{tabular}{cc}
\includegraphics[width=.47\textwidth]{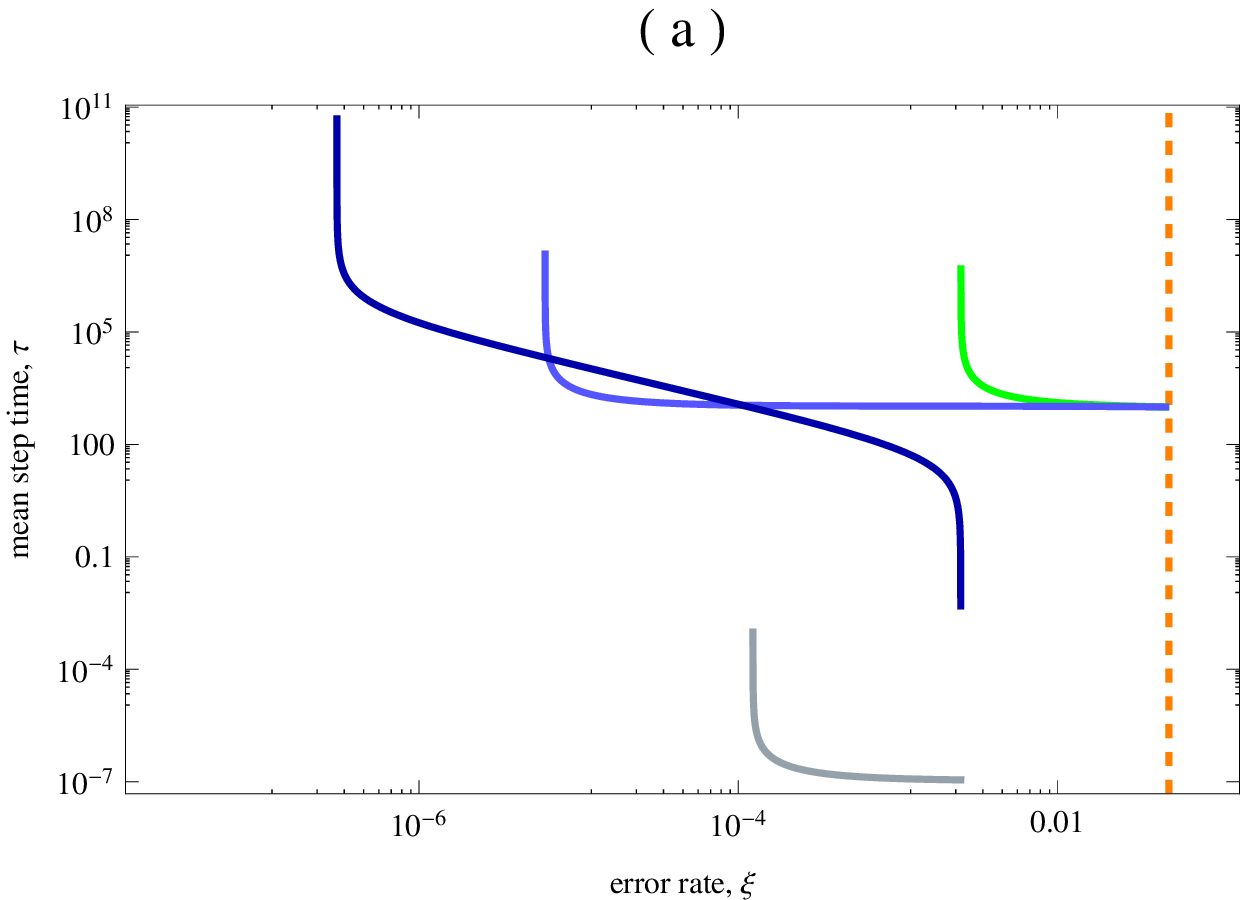} &
\includegraphics[width=.47\textwidth]{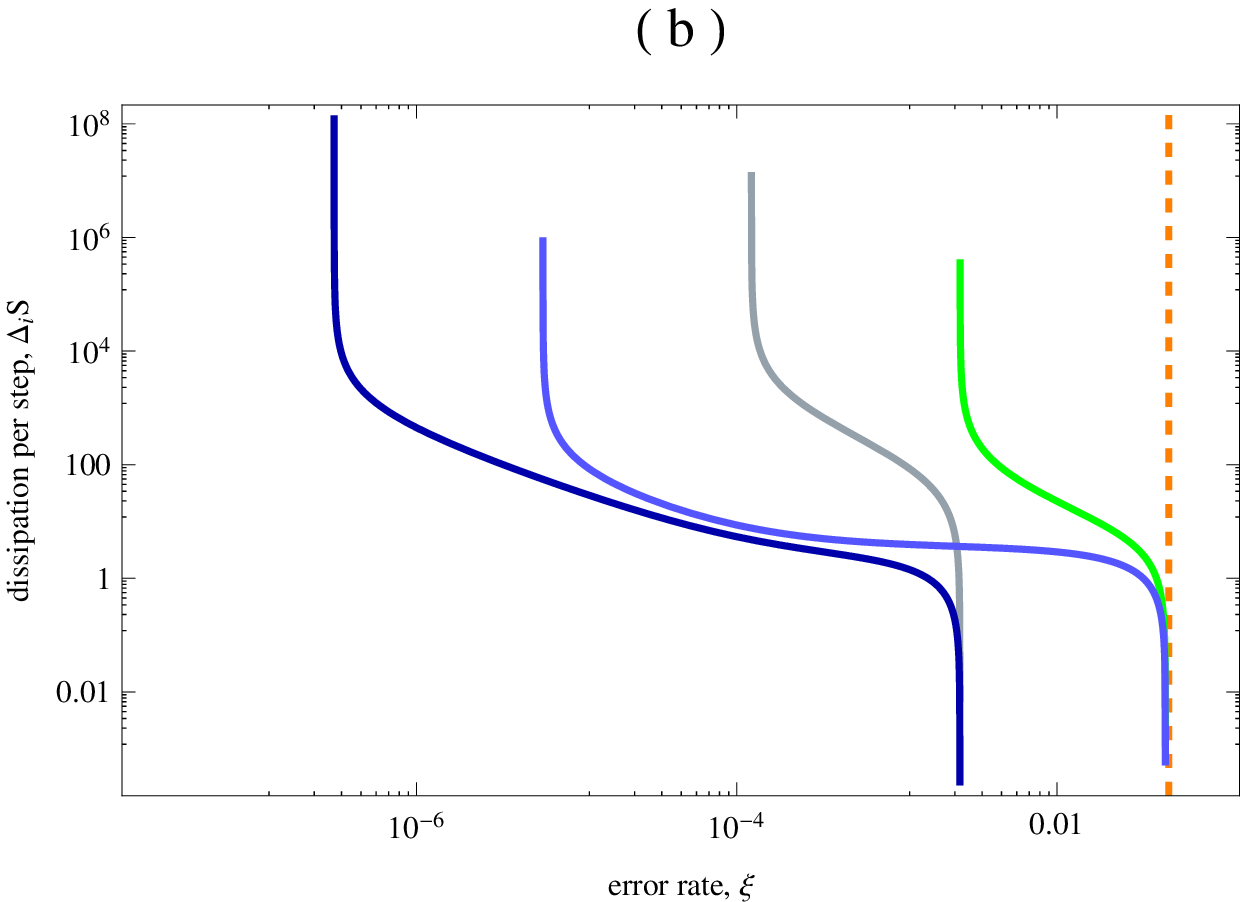} \\
\includegraphics[width=.47\textwidth]{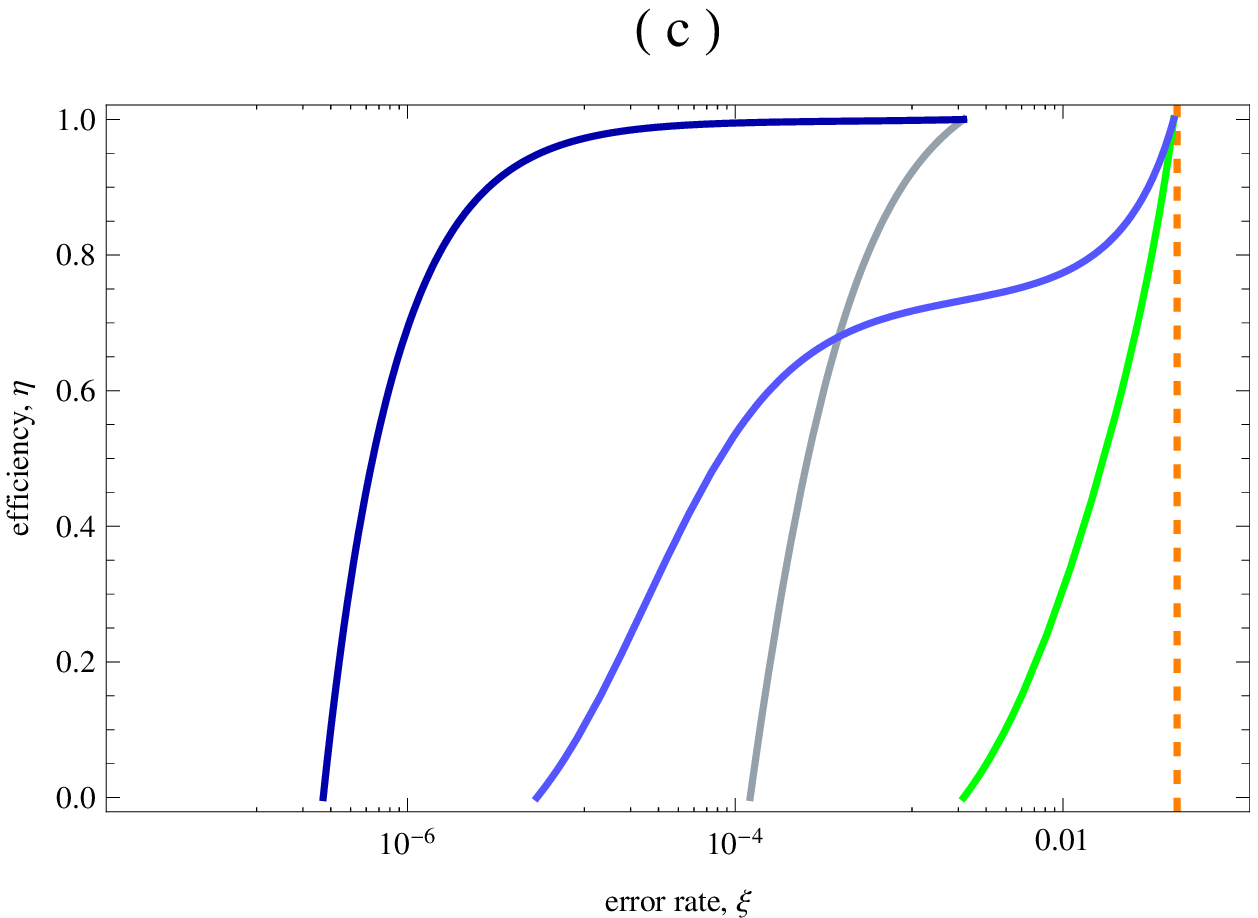}
\end{tabular}
\caption{
	Mean step time (a), dissipation per step (b), and efficiency (c) versus the error-rate for the Hopfield model.
	We have chosen the following values for the coefficients:
	$\epsilon = 10$, $\omega = 1$, $\gamma = 3$ and $(\delta, \delta_{\mathrm{p}}) = (0,0)$ (green curve: purely energetic discrimination), $(\delta, \delta_{\mathrm{p}}) = (6,0)$ (light blue curve: kinetic discrimination on the first chemical pathway), $(\delta, \delta_{\mathrm{p}}) = (0,6)$ (blue curve: kinetic discrimination on the proofreading chemical pathway), $(\delta, \delta_{\mathrm{p}}) = (6,6)$ (dark blue curve: kinetic discrimination on both chemical pathways).
	The other constant, namely $\omega_{\mathrm{i}}$, $\omega_{\mathrm{p}}$, $\epsilon_{\mathrm{i}}$, $\epsilon_{\mathrm{p}}$, are those which minimize the error-rate function $\xi$. 
	In this way, the minimum achievable error rate is recovered by the trade-offs.
	They have been obtained by numerically minimizing the error-rate function given the discrimination constants and the driving energy $\epsilon$, and are consistent with the values predicted by Hopfield in \cite{hopfield:proofreading}
	The dashed orange curve refers to the equilibrium error rate for $\gamma = 3$. 
	Therefore, in the Hopfield model one consistently obtains smaller error rates than in the MM model for equal values of the discrimination constants. 
	The light blue and blue curves highlight that kinetic and energetic discrimination regimes yield a resulting lower error rate only when they cooperate in the proofreading pathway.
	Remarkably, when the kinetic discrimination predominates on the first pathway the process becomes faster, more dissipative and more efficient.
	Finally, it is worth noting that high efficiency at lower error rates appears as a general trait of the kinetic discrimination for simple models, like the MM and the Hopfield ones.
}
\label{fig:h:tos}
\end{figure}
In the simple MM scheme, all the dissipated energy is spent to drive the process faster and not to proofread the outcome. However, this acceleration enhances the accuracy of the assembly only in the kinetic discrimination regime. In order to overstep the error rate thresholds obtained with the MM one needs a chemical mechanism in which the chemical energy provided by the environment is exploited to proofread the outcome, enhancing the pre-existing discrimination.

This can be realized by properly coupling the internal network directly with a chemical force. The first reaction scheme embedding this idea was independently proposed by J.~Ninio~\cite{ninio:proofreading} and by J.~J.~Hopfield~\cite{hopfield:proofreading}. We shall describe this model in the version proposed by Hopfield and shown in~fig.~\ref{fig:h:tos}. Two forces are coupled to the cycles of the network (one for each substrate) and drive the chemical complex preferentially in the direction $\mathsf{\&} \rightarrow \mathsf{\&} \mathsf{s}^{*} \rightarrow \mathsf{\&} \mathsf{s} \rightarrow \mathsf{\&}$. Indeed, with reference to the energy landscape shown in fig.~\ref{fig:h}(bottom) the system energy drops of an amount equal to $\mathcal{A} \equiv \epsilon + \epsilon_{\rmi} + \epsilon_{\mathrm{p}}$ when a cycle is performed. We are assuming that the chemical forces are equal for any substrate and the related affinity $\mathcal{A}$ is the same. The above-mentioned energy scheme traces the original Hopfield idea, but it is also provided with the kinetic discrimination scheme introduced in~\cite{bennett:proofreading} and successively developed in~\cite{sartori:discrimination}. According to Hopfield's idea, the first and the proofreading reactions discriminate. On the one hand, the discrimination can be kinetic with discrimination constants denoted by $\delta$ and $\delta_{\mathrm{p}}$, for the first and the proofreading reaction, respectively. On the other hand the discrimination can also be energetic if, e.g., the binding energy of the wrong pre-catalysis state, $\mathsf{\&} \mathsf{w}$, is larger than the corresponding one for the right pre-catalysis state---here $\gamma$ again denotes the difference of these energies. In this way, the wrong substrate is more likely to be rejected through the proofreading pathway.

This picture can be represented by the rate constants, which are given by equation~(\ref{expr:mm}) for the first reaction, and by
\begin{eqnarray}
\label{eq:h:rc}
\eqalign{
		 	h_{\mathsf{r}}^{+} = \omega_{i} \rme^{\epsilon_{i}},\qquad &h_{\mathsf{r}}^{-} = \omega_{i}, \\ 
		 	h_{\mathsf{w}}^{+} = \omega_{i} \rme^{\epsilon_{i}}, &h_{\mathsf{w}}^{-} = \omega_{i},}
		& \qquad
		\eqalign{
		 	K_{\mathsf{r}}^{+} = \omega_{p} \rme^{-\delta_{p}},	\qquad & K_{\mathsf{r}}^{-} = \omega_{p} \rme^{\epsilon_{p} - \delta_{p}}, \\ 
		 	K_{\mathsf{w}}^{+} = \omega_{p},					& K_{\mathsf{w}}^{-} = \omega_{p} \rme^{\epsilon_{p} + \gamma }.}
\end{eqnarray}
for the second and the proofreading reactions.
The related steady state can be evaluated, and the relevant kinetic and thermodynamic quantities derived.
Finally, inverting the error rate as a function of the catalytic rate we obtain the trade-offs, shown in fig.~\ref{fig:h:tos}. 

In our thermodynamic description, the relevant parameters are the discrimination constants ($\gamma, \delta, \delta_{\mathrm{p}}$), since they determine the operation regime of the proofreading mechanisms (a detailed discussion on the role of the other constants is discussed in~\cite{hopfield:proofreading}).
With respect to the MM scheme, the working regimes of the Hopfield model are characterized by lower error rates on equal values of the discrimination constants.
As already discussed by Hopfield, in the purely energetic discrimination regime the minimum achievable error rate equals the square of that obtained by the MM model (i.e., the equilibrium value) for the same values of the discrimination constants.
When also the kinetic discrimination is taken into account the lowest achievable error rate becomes
\begin{equation}
\label{expr:hopfield:minimum_er} 
	\xi_{\min} \simeq \rme^{ - \left( \max \left\{ \gamma, \delta \right\} + \gamma + \delta_{\mathrm{p}}\right) }.
\end{equation}
It is worth noting that thanks to the proofreading reaction pathway, the discrimination regimes can be partially mixed to further reduce the error rate, fig.~\ref{fig:h:tos}.
Indeed, although the discrimination factors of the first reaction pathway do not mix to reduce the error rate, those of the proofreading reaction do (blue curves in fig.~\ref{fig:h:tos}).
The apparent contradiction with the results of~\cite{sartori:discrimination} is due to a different design of the free enthalpy landscape.
The presence of the kinetic factor also speeds up the assembly at the price of a larger free-energy consumption (light blue curve in fig.~\ref{fig:h:tos}).
Interestingly, the process becomes more efficient, too (fig.~\ref{fig:h:tos}c).

Let us point out that the minimum error rate is always achieved in the vanishing catalysis rate limit.
In this regime all the free energy provided to the system is spent to proofread at the expense of the process speed.
Hence, both the mean step time and the dissipation diverge approaching the minimum error rate.

\section{The Murugan-Huse-Leibler (MHL) Model}
\label{sec:Murugan}
\begin{figure}[htb]
\centering
	\includegraphics[width=.9\textwidth]{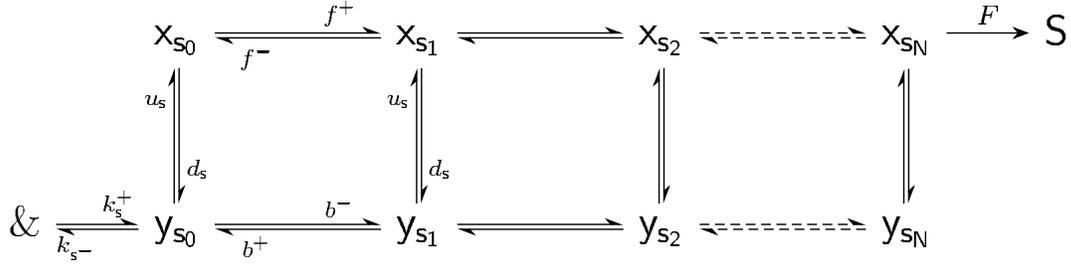}
\caption{MHL (``railroad'') model reaction scheme for the generic substrate $\mathsf{s}$.}
\label{fig:m:scheme}
\end{figure}

The idea of embedding the network with more forces in order to reduce the error rate has been pursued in~\cite{murugan:speed}, where a general scheme involving many reaction cycles was introduced. The authors of this work were able to show that the minimal error rate (with given value of the energetic discrimination constant $\gamma$) depends on the network topology, i.e., on the number of cycles appearing in the reaction graph. This class of~models was further explored in a more general setting in~\cite{murugan:regimes}. In particular, the ``railroad'' scheme shown in fig.~\ref{fig:m:scheme} was introduced, as the most efficient one among a large class of reactions. We shall refer to this model as the MHL (Murugan-Huse-Leibler) model.  Any substrate chemical network is coupled with $N$ chemical forces of equal affinity $\mathcal{A}$.
These forces act on the $N$ cycles of type $ y_{\mathsf{s}_{n}} \rightarrow x_{\mathsf{s}_{n}} \rightarrow x_{\mathsf{s}_{n+1}} \rightarrow y_{\mathsf{s}_{n+1}} \rightarrow y_{\mathsf{s}_{n}} $ and drive the chemical complex towards the final product state on the $x$-chain of reactions and towards the free enzyme state on the $y$-chain of reactions.
The $x_{\mathsf{s}_{i}} \rightleftharpoons y_{\mathsf{s}_{i}}$ reactions discriminate the substrates either kinetically or energetically.
The discrimination constants, denoted  by $\gamma$ and $\delta$, and will be taken to be for any of the $N+1$ pathways $x_{\mathsf{s}_{i}} \rightleftharpoons y_{\mathsf{s}_{i}}$.

\begin{figure}[htb]
\centering
\begin{tabular}{cc}
		{\includegraphics[width=.47\textwidth]{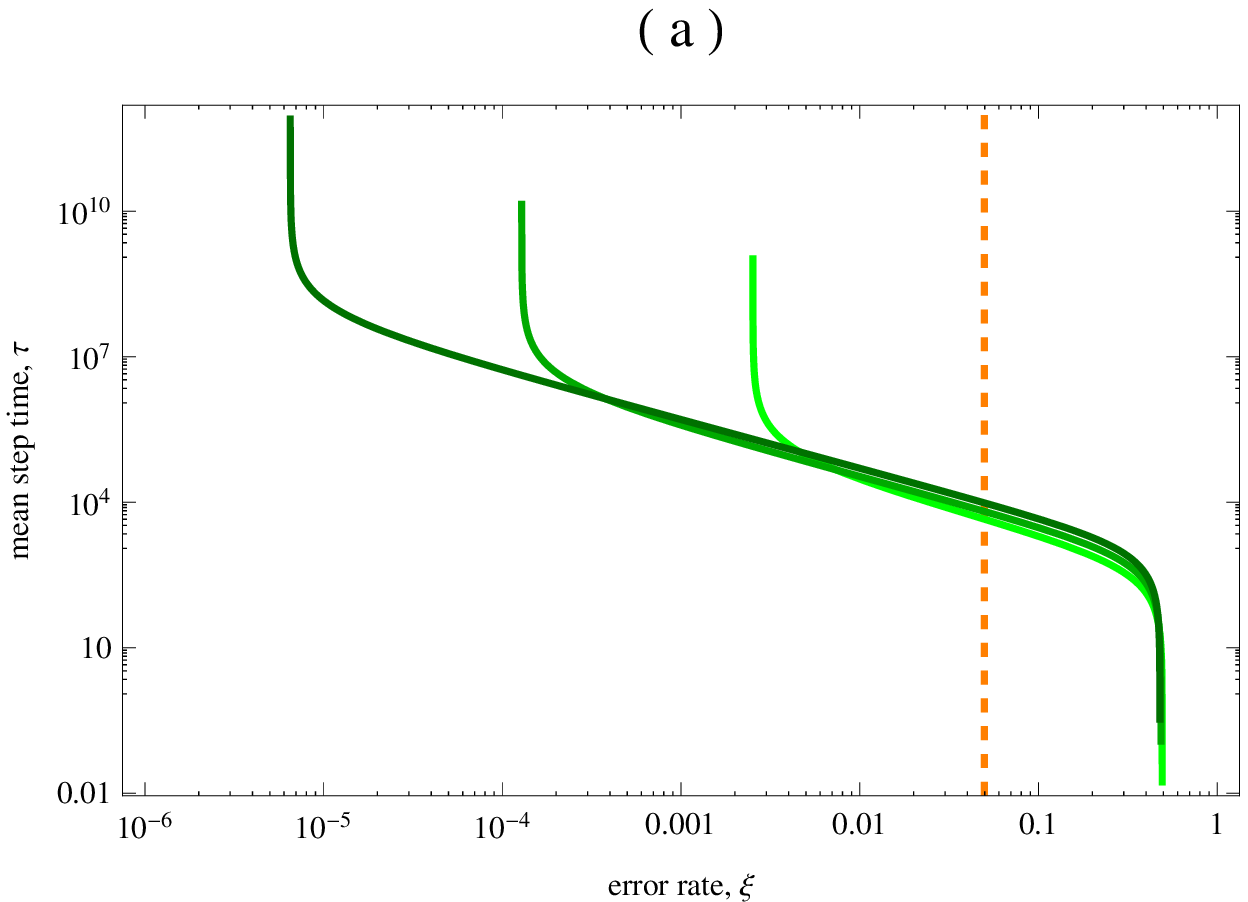}} &
		{\includegraphics[width=.47\textwidth]{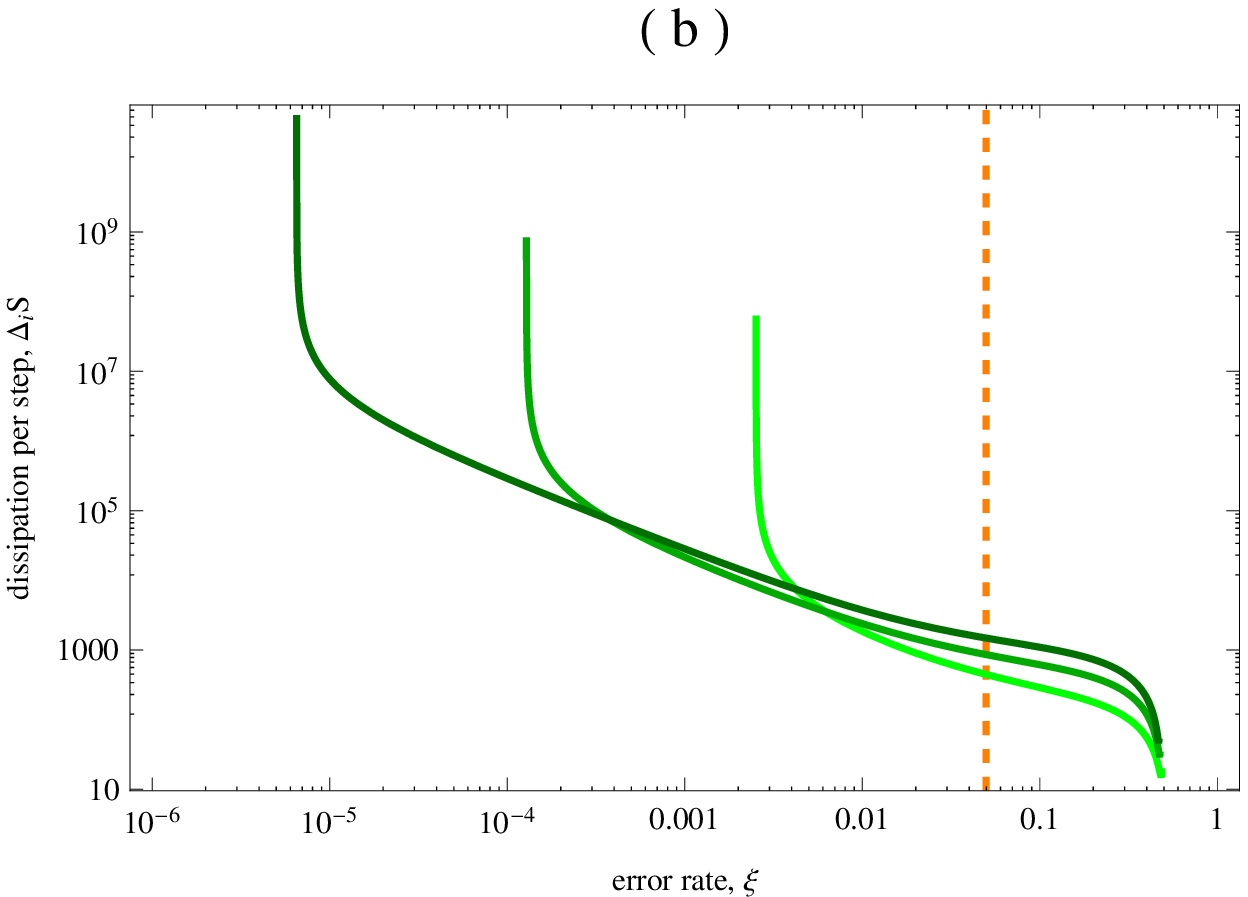}} \\
		{\includegraphics[width=.47\textwidth]{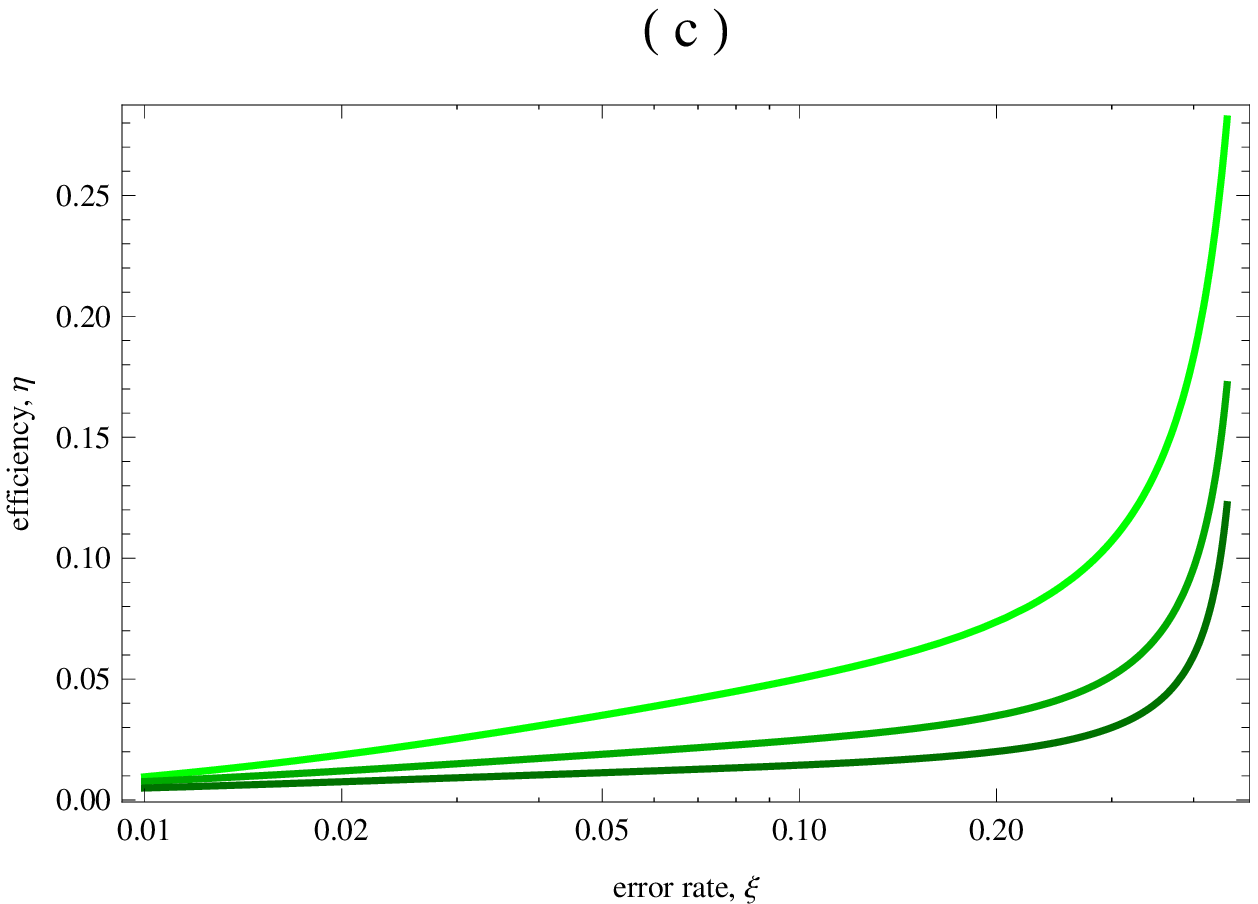}} &
		{\includegraphics[width=.47\textwidth]{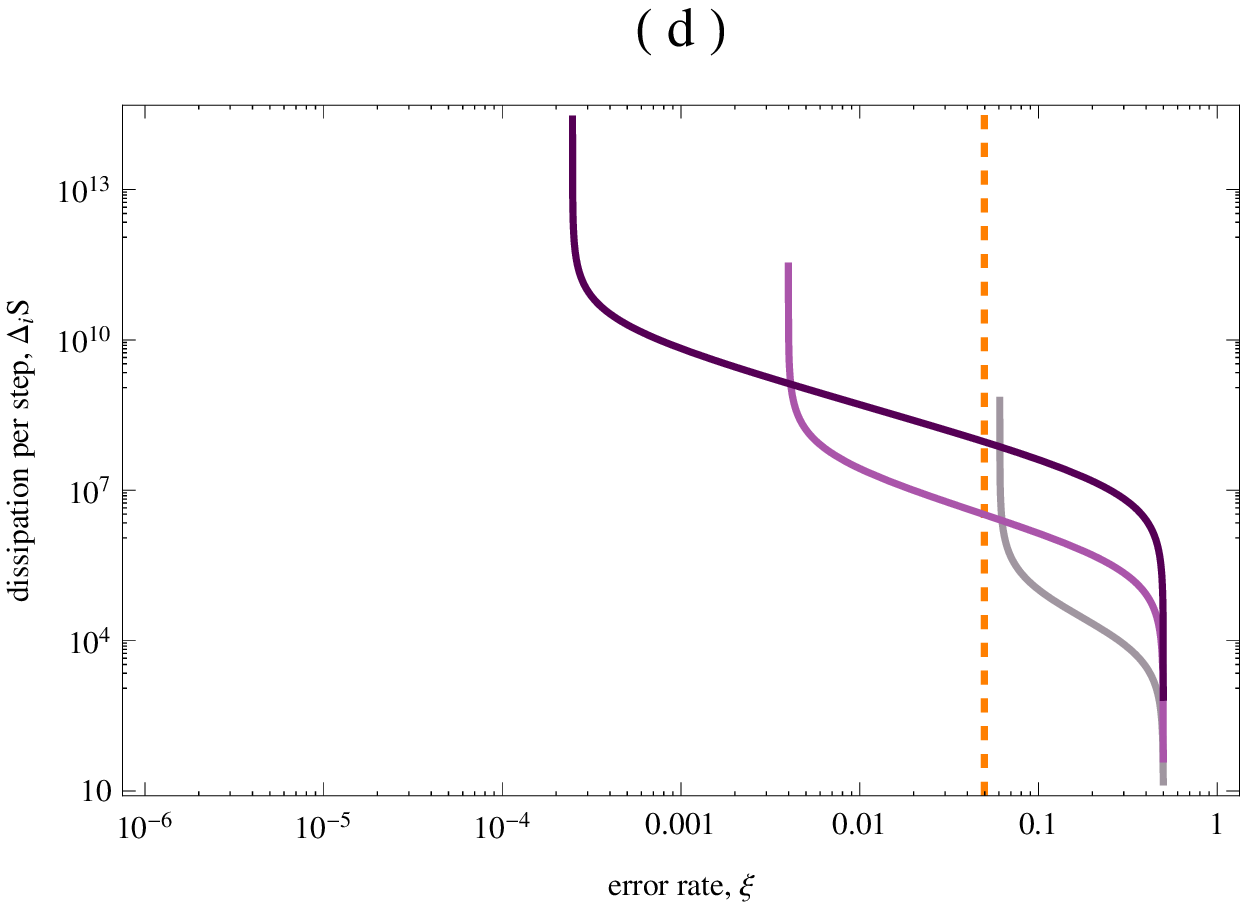}} 
\end{tabular}
\caption{
	Mean step time (a), dissipation per step (b), and efficiency (c) versus the error rate for the MHL model in the purely energetic discrimination regime ($\gamma = 3$ and $\delta = 0$).
	The plot (d), instead, shows the dissipation per step for the same model in the purely kinetic discrimination regime ($\gamma = 0$ and $\delta = 3$).
	The curves are distinguished by the number of cycles in the network: light green and light purple curves, $N = 1$; green and purple curves, $N = 2$; dark green and dark purple curves, $N = 3$.
	The curves in (a--c) correspond to the following values of the parameters: $\gamma = 3$, $\epsilon_{\U} = 8$, $\epsilon_{\F} = 8$ and $\epsilon_{\B} = 8$, which represent the driving energy related to the reactions $x_{s_{i}} \rightarrow y_{s_{i}}$, $x_{s_{i}} \rightarrow  x_{s_{i+1}}$ and $y_{s_{i+1}} \rightarrow  y_{s_{i}}$, respectively.
	The other constant, $\omega_{\U}$, $\omega_{\F}$, $\omega_{\B}$, are those which minimize the minimal error rate $\xi_{\min}$, when the other parameters are kept fixed.
	The first two plots in (a) and (b) highlight the lowering of the error rate as the number of forces increase.
	However, both the mean step time and the entropy production increase and the system exhibits a progressively lower efficiency (c).
	Interestingly, in the kinetic regime (d), the system needs at least two forces in order to reduce the error rate.
	Let us observe that the model embedding just one cycle is very similar to Hopfield scheme, except for the non-discriminating pathways $\mathsf{\&} \rightleftharpoons \mathsf{y_{s_0}}$.
	We thus suppose that these last pathways, detaching the free enzyme state from the part of the network performing the proofreading, spoil the discrimination in the kinetic regime.
	Finally, for the plot (d) the numeric constants chosen are: $\omega = 1$, $\epsilon = 10$, $\delta = 3$.
	The values of the other coefficients are those which minimize the minimal error rate $\xi_{\min}$, and are given by $\epsilon_{\U} = 8$, $\epsilon_{\F} = 9$, $\epsilon_{\B} = 9$, $\omega_{\U} = 0.1$, $\omega_{\F} = 10$, $\omega_{\B} = 10$).
	}
\label{fig:m:tos}
\end{figure}

We will focus our analysis to the behavior of the chemical network for varying number of independent forces, and thus of cycles.
For this purpose we neglect the discrimination performed on the first stage $\mathsf{\&} \rightleftharpoons y_{\mathsf{s}_{1}}$ and assume that the related rate constants are the same for both substrates: $k_{\mathsf{r}}^+ = k_{\mathsf{w}}^+ = \omega e^\epsilon$ and $k_{\mathsf{r}}^- = k_{\mathsf{w}}^- = \omega$.

Thus the rate constants related to the ladder part of the network are expressed by
\begin{equation}
\label{eq:m:rc}
\eqalign{
		 u_{\mathsf{r}} = \omega_{\U} \rme^{\epsilon_{U} + \delta},\qquad &d_{\mathsf{r}} = \omega_{u} \rme^{\delta}, \\ 
		 u_{\mathsf{w}} = \omega_{\U} \rme^{\epsilon_{u}}, &d_{\mathsf{w}} = \omega_{u} \rme^{\gamma},}
		 \qquad
		\eqalign{
		 	f^{+} = \omega_{\F},\qquad	& f^{-} = \omega_{\F} \rme^{-\epsilon_{\F}}, \\ 
		 	b^{+} = \omega_{\B},	& b^{-} = \omega_{\B} \rme^{-\epsilon_{\B}}.}
\end{equation}

We can evaluate the error rate, the mean step rate and the efficiency by solving the steady-state dynamics, inverting the related error-rate function, and substituting the result in the relevant expressions, obtaining the results shown in fig.~\ref{fig:m:tos}.

The relevant parameters are again the discrimination constants. In order to simplify the analysis, we discuss the energetic and kinetic discrimination regimes separately.
The trade-offs in the energetic regime, taking into account different numbers of cycles, are shown in the figures \ref{fig:m:tos} (a) and~(b).
In agreement with~\cite{murugan:speed}, the minimum error rate decreases exponentially as the number of cycles increases ($\xi_{\mathrm{min}} \sim \rme^{-(N+1)\gamma}$).
However, when the number of cycles increases, both the mean step time and the dissipation per step increase, for equal values of the discrimination constants and error rate.
This also plays a role on the efficiency, which accordingly decreases, fig.~\ref{fig:m:tos}(c).

An analogous scenario appears in kinetic discrimination regime, whose dissipation-error trade-off is shown in fig.~\ref{fig:m:tos}(d).
The network needs at least two cycles to achieve error rates lower than that of the MM model, for equal values of the constants ($\xi_{\mathrm{min}} \sim \rme^{-N\delta}$).
We thus infer that the most effective proofreading action is thus the result of the optimal combination of the network topology~and discrimination regime.

\section{Michaelis-Menten model with correlations between consecutive steps}
\label{sec:correlations} 

As mentioned in sec.~\ref{sec:description}, in all the enzyme-assisted assembly schemes described so far all steps are similar to and independent from one another. In this section we take into account a simple scheme in which consecutive steps are correlated.

\begin{figure}
\centering
	\includegraphics[width=.7\textwidth]{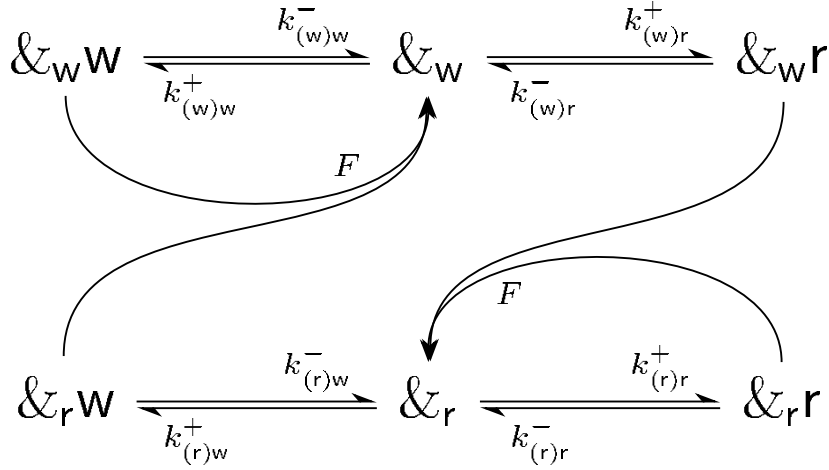}
\caption{
	Michaelis-Menten (MM) model with correlations between consecutive steps.
}
\label{fig:corr:scheme}
\end{figure}

We assume that a wrong step affects the system discriminatory power, i.e., that when the last step produces the wrong outcome, the discrimination constants decrease. We introduce this simple correlation scheme in the MM model and obtain the chemical network shown in fig.~\ref{fig:corr:scheme}. Here, we consider two parallel reaction schemes: the first describes the steps following an error (denoted by the label $\mathsf{w}$ on the enzyme state, $\mathsf{\&}_{\mathsf{w}}$), while the second describes the steps following a correctly terminated step (denoted by the label $\mathsf{r}$ on the enzyme state, $\mathsf{\&}_{\mathsf{r}}$). Finally, the connection between these two reaction networks is realized by the catalysis transitions. In this way, when the enzyme catalyzes the final reaction of the wrong substrate, the system comes back to the ``wrong" free enzyme state. Instead, when the enzyme catalyzes the final reaction of the right substrate, the system comes back to the ``right" free enzyme state. As for the MM model, we assume that the catalysis transition does not discriminate and thus that the catalytic rate $F$ is the same for the four possible outcomes.

\begin{figure}
\centering
	\begin{tabular}{cc}
		{\includegraphics[width=.47\textwidth]{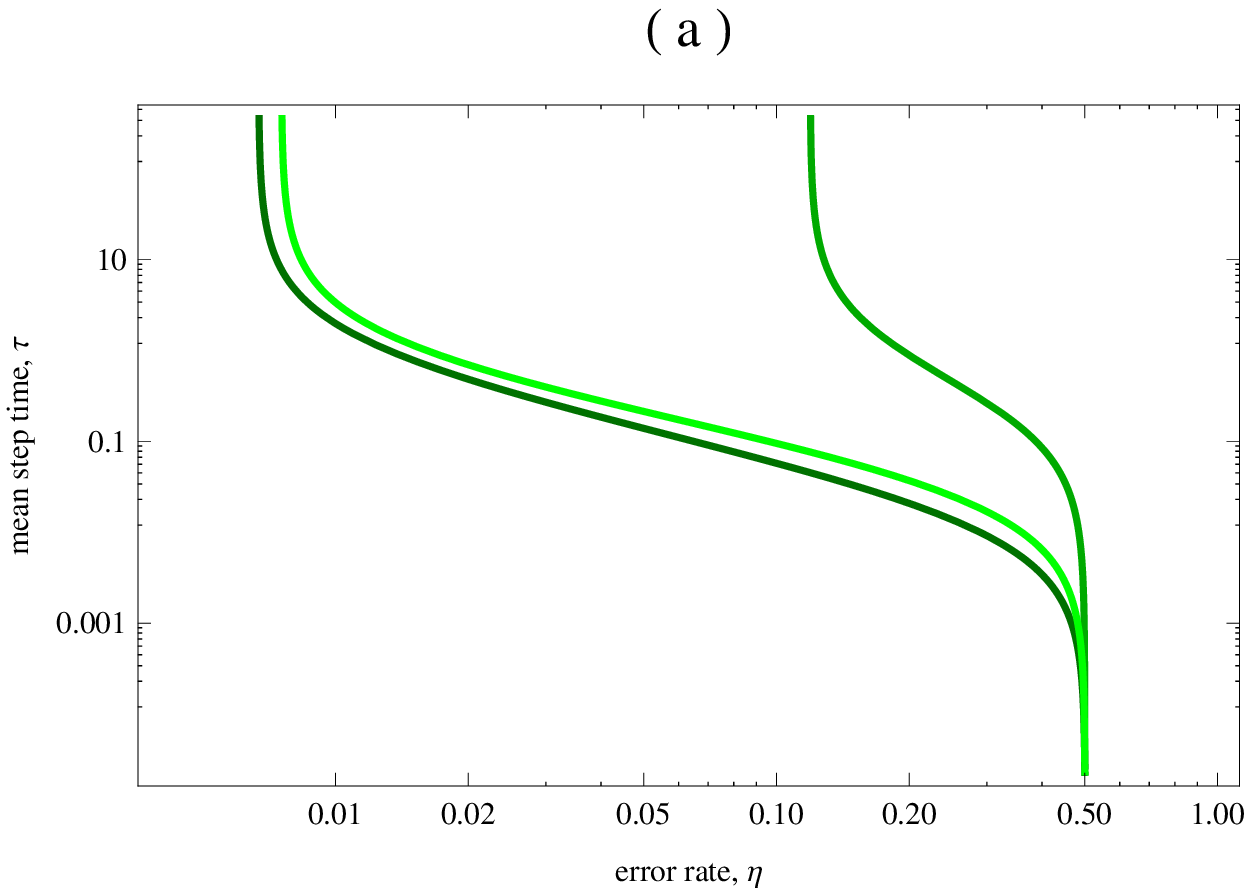}} \quad &
		{\includegraphics[width=.47\textwidth]{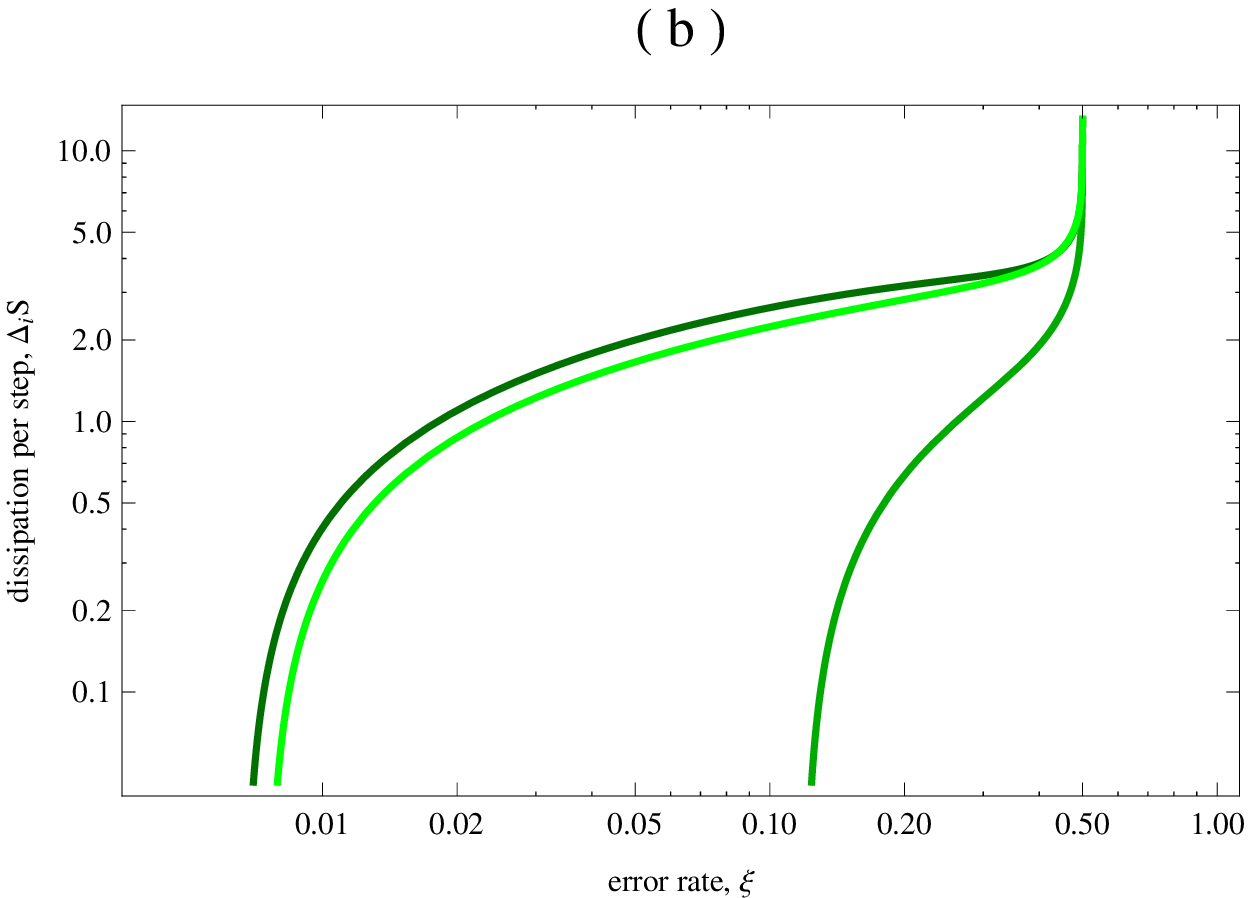}} \\
		{\includegraphics[width=.47\textwidth]{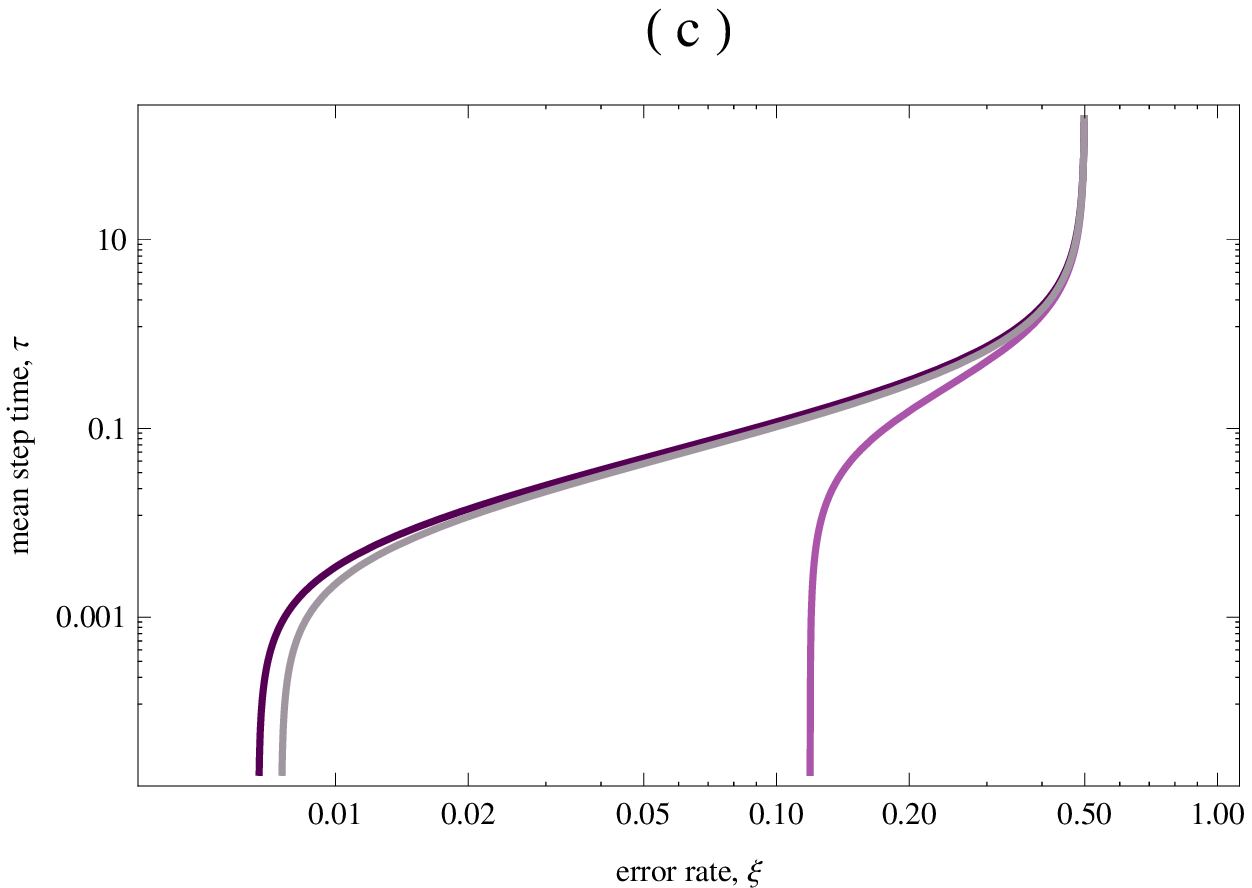}} &
		{\includegraphics[width=.47\textwidth]{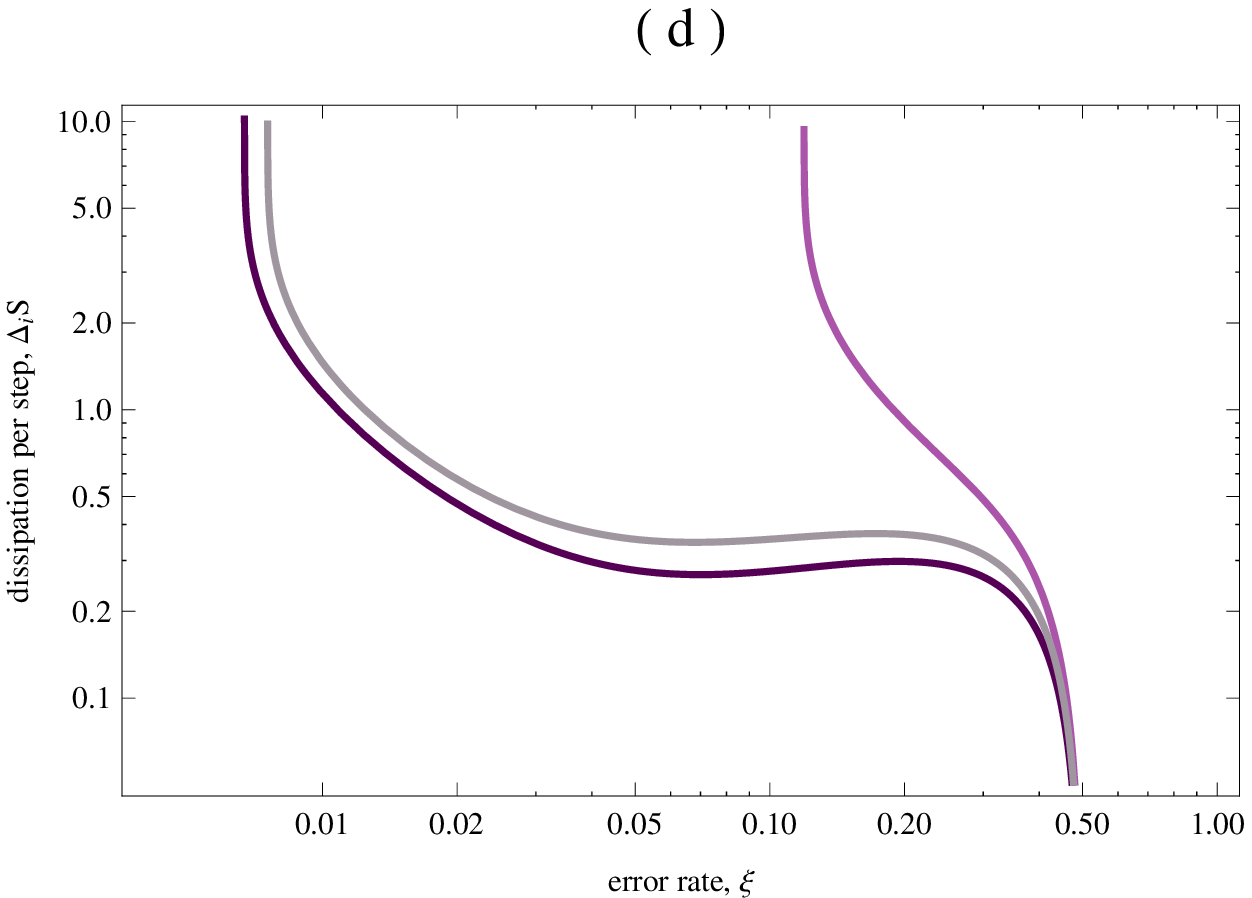}}
	\end{tabular}
\caption{
	The light green and light purple lines represent the mean step time (a, c), and entropy production (b, d), as a function of the error rate for the MM model with correlations between consecutive steps.
	(a) and (b) correspond to the energetic discrimination regime $\gamma_{\mathsf{w}} = 2, \gamma_{\mathsf{r}} = 5$, while (c) and (d) correspond to the kinetic one $\delta_{\mathsf{w}} = 2, \delta_{\mathsf{r}} = 5$.
	The curves are compared with the trade-offs obtained with the MM model (darker lines) for the same values of $\epsilon$ and $\omega$, but taking into account separately the discrimination constants used for the model with correlations.
	We have chosen the following values for the driving force and the overall rate scale: $\epsilon = 10$ and $\omega = 1$ (let us remind that these values are by assumption common for both the ``right'' and the ``wrong'' network).
	Remarkably, the reduction of the network discriminatory power is small.
	Indeed, the trade-offs we obtain are close to those obtained in the MM model by considering just the discrimination constants on the ``right'' pathway, namely the darkest curves (with $\gamma = \gamma_{\mathsf{r}}$ and $\delta = \delta_{\mathsf{r}}$).
}
\label{fig:corr:tofs}
\end{figure}

The type of correlation described above is implemented in the rate constants. For both networks they have the same expressions as in (\ref{expr:mm}) but with different discrimination constants. Following our hypothesis, the discrimination constants related to the wrong network, $\gamma_{\mathsf{w}}$ and $\delta_{{\mathsf{w}}}$, are smaller than those for the right network, $\gamma_{\mathsf{r}}$ and~$\delta_{\mathsf{r}}$.

We can solve the master equation in the steady-state regime and evaluate all the relevant quantities. The error rate and the mean step time are expressed by
\begin{equation}
\eqalign{
	\xi \equiv \frac{ \bar{p}_{(\mathsf{w})\mathsf{w}} + \bar{p}_{(\mathsf{r})\mathsf{w}} }{ \bar{p}_{(\mathsf{w})\mathsf{w}} + \bar{p}_{(\mathsf{w})\mathsf{r}} + \bar{p}_{(\mathsf{r})\mathsf{w}} + \bar{p}_{(\mathsf{r})\mathsf{r}}  } ,\label{expr:corr:er} \\
	\tau \equiv \frac{ 1 }{ F  \left(\bar{p}_{(\mathsf{w})\mathsf{w}} + \bar{p}_{(\mathsf{w})\mathsf{r}} + \bar{p}_{(\mathsf{r})\mathsf{w}} + \bar{p}_{(\mathsf{r})\mathsf{r}}\right) .}  \label{expr:corr:mst}
}
\end{equation}
The thermodynamic quantities can be expressed as in (\ref{def:epps}) and~(\ref{def:effps}), taking care of excluding the catalysis transitions from the sums.
However, we cannot express the error rate as a function of the catalytic rate in the general case (with four discrimination parameters).
We will thus limit our analysis to the cases of purely energetic and purely kinetic discrimination, in which this expression is possible.

Typical trade-offs are plotted in fig.~\ref{fig:corr:tofs} together with the trade-offs obtained by the MM with discrimination constants equal to 
$\gamma_{\mathsf{w}}$ and $\gamma_{\mathsf{r}}$, and with $\delta_{\mathsf{w}}$ and $\delta_{\mathsf{r}}$.
While the expression for the minimum achievable error rates, are given by
\begin{equation}
\eqalign{
	\xi_{\min} &= \frac{ \rme^{\gamma_\mathsf{w}} + 1 }{ \rme^{\gamma_\mathsf{r}+\gamma_\mathsf{w}} + 2 \rme^{\gamma_\mathsf{w}} + 1} \simeq  \left( \frac{1 + \rme^{\gamma_{\mathsf{w}}}}{\rme^{\gamma_\mathsf{w}}} \right) \rme^{-\gamma_{\mathsf{r}}} ,\\
	\xi_{\min} &= \frac{ \rme^{\delta_\mathsf{w}} + 1 }{ \rme^{\delta_\mathsf{r}+\delta_\mathsf{w}} + 2 \rme^{\delta_\mathsf{w}} + 1} \simeq  \left( \frac{1 + \rme^{\delta_{\mathsf{w}}}}{\rme^{\delta_\mathsf{w}}} \right) \rme^{-\delta_{\mathsf{r}}} ,
}
\end{equation}
in the energetic and kinetic regimes, respectively.

It is clear from this result that the discrimination loss due to a wrong outcome only slightly influences the global behavior of the kinetics.
Indeed, for high values of $\gamma_{\mathsf{r}}$ and $\delta_{\mathsf{r}}$ the difference between the nearest-neighbor and the simple MM model becomes negligible.

Let us observe that when proofreading mechanisms are taken into account, like in Hopfield model, the probability that the step ends with a wrong incorporation decreases with respect to the MM model. Thus we expect that in the Hopfield model with nearest-neighbor interactions, the discrimination loss due to a wrong monomer incorporation would influence its global behavior even less than in the MM model.
We conclude by remarking that these results are not in contradiction with the effective role of correlations in biological systems.
As already mentioned, in template-assisted assembly processes, correlations play a role in the individuation and proofreading of error, i.e., they can influence the internal rate constants when interaction with multiple substrates is taken into account.
\section{Conclusions}
\label{sec:Conclusions}
Natural biological processes, like the enzyme assisted assembly processes that we have considered, must satisfy several, sometimes contradictory, requirements. The error rate should be lower than a threshold, the average duration of the process should not be too long, and the free-energy consumption should not be too large. It is likely that the detailed mechanisms of naturally occurring processes have evolved to optimize some combination of these quantities, whose expression depends on the requirements of the process itself. The method discussed in the present work, and the measure of efficiency that we have derived, can be useful in the evaluation of the performance of these mechanisms.

We have been able to investigate a situation in which successive completed processes are not fully independent. In our situation, we have found that the correlations do not significantly modify the performance of the mechanism. We can however envisage more general situations, in which hidden correlations show up in the system performance. We have studied in a simple case the interplay of kinetic and energetic discrimination in many-cycle proofreading mechanisms, generalizing the results of~\cite{murugan:speed,murugan:regimes}. We have identified a regime in simple proofreading models in which the introduction of a kinetic discrimination step improves both the speed and the efficiency of the process for a given discrimination rate. We can speculate that naturally occurring mechanisms work in a regime that optimizes some combination of these quantities---the particular combination depending on the details of the role of the process in the global cellular biochemical network.

We can envisage extending the research to more general models, in which also the external chemical species can occur with varying concentration, driving the system out of equilibrium in several modes, as suggested, e.g., in~\cite{polettini:networks1}. This may lead to the understanding of more complex proofreading mechanisms, like those that take place in metabolic processes (cf.~\cite{linster:metabolic}). Indeed, the recent advances in the theory of the interplay of information and thermodynamics in non-equilibrium processes can help us in reaching a more precise understanding on the constraint obeyed by these fundamental processes in living systems.

\ack 
The authors are grateful to M. Esposito for encouragement and illuminating discussions. They warmly thank D.~Huse, S.~Leibler, A.~Murugan, S.~Pigolotti and M.~Polettini for suggestions and feedback.

\section*{References}
\bibliography{bib_database}

\providecommand{\newblock}{}
\begin{thebibliography}{10}
\expandafter\ifx\csname url\endcsname\relax
  \def\url#1{{\tt #1}}\fi
\expandafter\ifx\csname urlprefix\endcsname\relax\def\urlprefix{URL }\fi
\providecommand{\eprint}[2][]{\url{#2}}

\bibitem{ninio:proofreading}
Ninio J 1975 {\em Biochimie\/} {\bf 57} 587--595

\bibitem{hopfield:proofreading}
Hopfield J~J 1974 {\em Proc. Natl. Acad. Sci. U.S.A.\/} {\bf 71} 4135--4139

\bibitem{landauer:irreversibility}
Landauer R 1961 {\em IBM J. Res. Dev.\/} {\bf 5} 183--191

\bibitem{bennett:notes}
Bennett C~H 2003 {\em Stud. Hist. Philos. M. P.\/} {\bf 11} 501--510

\bibitem{esposito:farfrom}
Esposito M and Van~den Broeck C 2011 {\em Europhys. Lett.\/} {\bf 95} 40004

\bibitem{hopfield:evidence}
Hopfield J~J, Yamane T, Yue V and Coutts S~M 1976 {\em Proc. Natl. Acad. Sci.
  U.S.A.\/} {\bf 73} 1164--1168

\bibitem{yamane:evidence}
Yamane T and Hopfield J 1977 {\em Proc. Natl. Acad. Sci. U.S.A.\/} {\bf 74}
  2246--2250

\bibitem{bennett:proofreading}
Bennett C~H 1979 {\em BioSystems\/} {\bf 11} 85--91

\bibitem{andrieux:randomness}
Andrieux D and Gaspard P 2008 {\em Europhys. Lett.\/} {\bf 81} 28004

\bibitem{andrieux:copol08}
Andrieux D and Gaspard P 2008 {\em Proc. Natl. Acad. Sci. U.S.A.\/} {\bf 105}
  9516--9521

\bibitem{andrieux:erasure}
Andrieux D and Gaspard P 2013 {\em Europhys. Lett.\/} {\bf 103} 30004

\bibitem{arias:DNAentropy}
Arias-Gonzalez J~R 2012 {\em PLoS ONE\/} {\bf 7} e42272

\bibitem{voliotis:DNAtranscription}
Voliotis M, Cohen N, Molina-París C and Liverpool T~B 2008 {\em Biophys. J.\/}
  {\bf 94} 334--348

\bibitem{murugan:speed}
Murugan A, Huse D~H and Leibler S 2012 {\em Proc. Natl. Acad. Sci. U.S.A.\/}
  {\bf 109} 12034

\bibitem{murugan:regimes}
Murugan A, Huse D~H and Leibler S 2014 {\em Phys. Rev. X\/} {\bf 4} 021016

\bibitem{sartori:discrimination}
Sartori P and Pigolotti S 2013 {\em Phys. Rev. Lett.\/} {\bf 110} 188101

\bibitem{parrondo:thermodynamics}
Parrondo J~M~R, Horowitz J~M and Sagawa T 2015 {\em Nature Phys.\/} {\bf 11}
  131–139

\bibitem{celani:harvesting}
{Bo} S, {Del Giudice} M and {Celani} A 2015 {\em J. Stat. Mech. Theor. Exp.\/}
  {\bf 1} 01014 (\textit{Preprint} \eprint{1408.5128})

\bibitem{seifert:efficiency}
Barato A~C, Hartich D and Seifert U 2014 {\em New J. Phys.\/} {\bf 16} 103024

\bibitem{seifert:analogy}
{Hartich} D, {Barato} A~C and {Seifert} U 2015 {\em ArXiv e-prints\/}
  (\textit{Preprint} \eprint{1502.02594})

\bibitem{alberts:cell5}
Alberts B, Johnson A, Lewis J, Raff M, Roberts K and Walter P 2007 {\em
  Molecular Biology of the Cell\/} 5th ed (Garland Science)

\bibitem{klamt:hypergraphs}
Klamt S, Haus U~U and Theis F 2009 {\em PLoS Comput. Biol.\/} {\bf 5}(5)
  e1000385

\bibitem{polettini:networks1}
Polettini M and Esposito M 2014 {\em J. Chem. Phys.\/} {\bf 141} 024117

\bibitem{hill:biochemical}
Hill T~L 2005 {\em Free Energy Transduction and Biochemical Cycle Kinetics\/}
  (Dover)

\bibitem{schnakenberg:network}
Schnakenberg J 1976 {\em Rev. Mod. Phys.\/} {\bf 48} 571--585

\bibitem{santalucia:DNAmotifs}
SantaLucia J and Hicks D 2004 {\em Annu. Rev. Biophys. Biomol. Struct.\/} {\bf
  33} 415--440

\bibitem{johnson:mismatch}
Johnson S and Beese L 2004 {\em Cell\/} {\bf 116} 803--816

\bibitem{hopfield80}
Hopfield J~J 1980 {\em Proc. Natl. Acad. Sci. U.S.A.\/} {\bf 77} 5248--5252

\bibitem{bialek:biophysics}
Bialek W 2012 {\em Biophysics: Searching for Principles\/} (Princeton
  University Press)

\bibitem{kampen:stochastic}
van Kampen N~G 2007 {\em Stochastic Processes in Physics and Chemistry\/} 3rd
  ed (Elsevier)

\bibitem{redner:first-passage}
Redner S 2001 {\em A guide to first-passage processes\/} (Cambridge University
  Press)

\bibitem{prigogine:thermodynamics}
Kondepudi D and Prigogine I 2015 {\em Modern Thermodynamics\/} 2nd ed (Wiley)

\bibitem{diana:erasing}
Diana G, Bagci G~B and Esposito M 2013 {\em Phys. Rev. E\/} {\bf 87}(1) 012111

\bibitem{andrieux:copol09}
Andrieux D and Gaspard P 2009 {\em J. Chem. Phys.\/} {\bf 130} 014901

\bibitem{shiraishi:masked}
Shiraishi N and Sagawa T 2015 {\em Phys. Rev. E\/} {\bf 91}(1) 012130

\bibitem{murashita14}
Murashita Y, Funo K and Ueda M 2014 {\em Phys. Rev. E\/} {\bf 90}(4) 042110

\bibitem{linster:metabolic}
Linster C~L, Van~Schaftingen E and Hanson A~D 2013 {\em Nat. Chem. Biol.\/}
  {\bf 9} 72–80

\end{thebibliography}

\end{document}